\documentclass[a4paper,10pt,twocolumn]{article}
\usepackage{amsmath}
\usepackage{graphicx}
\usepackage[authoryear]{natbib}
\usepackage{xcolor}
\usepackage[expproduct=times]{siunitx} 
\usepackage{url}
\usepackage[colorlinks=true,linkcolor=blue,citecolor=blue]{hyperref}
\usepackage[charter]{mathdesign} 
\usepackage[font=footnotesize,labelfont=bf,labelsep=endash,margin=0pt]{caption} 
\usepackage[title,header]{appendix}

\newlength{\myleftmargin} 
\newlength{\mytopmargin} 
\newlength{\myrightmargin} 
\newlength{\mybottommargin} 
\newcommand{\ocm}{\ensuremath{\text{OC}_\text{m}}}
\newcommand{\oca}{\ensuremath{\text{OC}_\text{a}}}
\newcommand{\oc}{\ensuremath{\text{OC}}}
\newcommand{\bone}{\ensuremath{\text{BONE}}}
\newcommand{\bv}{\ensuremath{\text{BV}}}
\newcommand{\bmu}{\ensuremath{\text{BMU}}}
\newcommand{\rank}{\ensuremath{\text{RANK}}}
\newcommand{\rankl}{\ensuremath{\text{RANKL}}}
\newcommand{\mcsf}{\ensuremath{\text{MCSF}}}

\newcommand{\mmp}{\ensuremath{\text{MMP}}}
\newcommand{\tauclean}{\ensuremath{\tau_\text{inhib}}}
\newcommand{\um}{\ensuremath{\micro\metre}}
\newcommand{\da}{\ensuremath{\text{day}}}
\newcommand{\days}{\ensuremath{\text{days}}}
\renewcommand{\b}[1]{{\boldsymbol{#1}}} 

\newcounter{modelstudy}
\def\m#1{\setcounter{modelstudy}{#1}Simulation~\Roman{modelstudy}}

\begin{document}

\title{{\bf Investigation of bone resorption within a cortical basic multicellular unit using a lattice-based computational model}}

\author{P R Buenzli$^\text{a}$, J Jeon$^{\text{b},1}$, P Pivonka$^\text{a}$, D W Smith$^\text{a}$ and P T Cummings$^\text{b,c}$}
\date{\small \vspace{-2mm}$^\text{a}$Engineering Computational Biology Group, FECM, The University of Western Australia, WA 6009, Australia
    \\$^\text{b}$Deptartment of Chemical and Biomolecular Engineering, Vanderbilt University, Nashville, TN 37235, USA
    \\$^\text{c}$Center for Nanophase Materials Sciences, Oak Ridge National Laboratory, Oak Ridge, TN 37831, USA
    \\\vskip 1mm \normalsize October 14, 2011\vspace*{-5mm}}
\twocolumn[
    \vskip-5mm
    \begin{@twocolumnfalse}
        \maketitle
        \begin{abstract}
            In this paper we develop a lattice-based computational model focused on bone resorption by osteoclasts in a single cortical basic multicellular unit (\bmu). Our model takes into account the interaction of osteoclasts with the bone matrix, the interaction of osteoclasts with each other, the generation of osteoclasts from a growing blood vessel, and the renewal of osteoclast nuclei by cell fusion. All these features are shown to strongly influence the geometrical properties of the developing resorption cavity including its size, shape and progression rate, and are also shown to influence the distribution, resorption pattern and trajectories of individual osteoclasts within the \bmu. We demonstrate that for certain parameter combinations, resorption cavity shapes can be recovered from the computational model that closely resemble resorption cavity shapes observed from microCT imaging of human cortical bone.
            
            \vspace{2mm}\noindent\textbf{Keywords:} bone resorption cavity, cortical Bone Multicellular Units (BMU), osteoclast--osteoclast interaction, osteoclast--bone interaction, osteoclast fusion, computational model
        \end{abstract}
        \vspace{5mm}
    \end{@twocolumnfalse}
    ]%
\protect\footnotetext[1]{Corresponding author
\\\textit{Email addresses:}
\\\texttt{pascal.buenzli@uwa.edu.au (P~R~Buenzli),
\\\texttt{junhwan.jeon@vanderbilt.edu} (J~Jeon),
\\\texttt{peter.pivonka@uwa.edu.au} (P~Pivonka),
\\\texttt{david.smith@uwa.edu.au} (D~W~Smith),
\\\texttt{peter.cummings@vanderbilt.edu} (P~T~Cummings)}}

\section{Introduction\label{introduction}}
The functional unit of cells in bone remodelling is the `basic multicellular unit' or \bmu\ \citep{frost1969,parfitt1983,parfitt-etal1987}. \bmu s are transient functional grouping of cells that progress through the bone, removing old bone and replacing it with new bone. A single \bmu\ comprises active multinucleated osteoclasts resorbing bone matrix at the front of the \bmu, and active osteoblasts towards the rear of the \bmu\ forming osteoid, which is later mineralized to form new bone matrix. As the \bmu\ progresses, the resorbing osteoclasts open-up a void space in the bone matrix called a \bmu\ `cutting cone', while the osteoblasts reduce the void space as bone is formed in a so-called \bmu\ `closing cone' \citep{parfitt1994}.

This paper focuses on a computational model of one aspect of cortical \bmu\ function, namely, osteoclasts opening up the cutting cone at the front of a single cortical \bmu. This resorption process determines in particular the diameter and morphology of the secondary osteon created by the \bmu. Here we show that for certain parameter combinations, we can demonstrate good agreement between the cutting cone shapes and osteon morphologies generated by the computational model, and the cutting cone shapes and osteon morphologies observed experimentally.

The development of our computational model relies upon detailed quantitative experimental observations. Much of what is known about the behaviour of cortical \bmu s has been obtained from quantitative analysis of two dimensional histomorphometric data. Indeed, quantitative analysis of \bmu s using stereology has been a primary focus of \cite{frost1969,frost1983}, \cite{parfitt1979,parfitt1983} and \cite{martin1994}. As a result of the research based on histomorphometric data, we now have reasonably reliable quantitative estimates for resorption properties, and we can define the general properties of a hypothetical `average cortical BMU'. For example, quantitative estimates have been made of resorption properties including osteoclast number, osteon diameter, and the duration and speed of movement of the \bmu\ through cortical bone \citep{parfitt1994,robling-castillo-turner}.

This histomorphometric data is supplemented by \textit{in vivo} data, which has given additional insights into osteoclast behaviour at the front of the \bmu. Most importantly, auto-radiographic studies of cell migration in cortical \bmu s have revealed that thymidine-labelled nuclei from osteoclast precursor cells fuse randomly with existing osteoclasts \citep{jaworski-duck-sekaly,miller,parfitt1994,parfitt-etal1996}. This fusion process leads to the continual renewal of the nuclei within the active osteoclasts at the front of \bmu s, which is believed to account for an increase in the osteoclasts' lifespan and/or the persistence of their resorptive activity.

This histomorphetric data and \textit{in vivo} data are supplemented by \textit{in vitro} cell-culture data, which has given additional insights into osteoclast behaviour. For example, the microscopic bone dissolution process operated by active osteoclasts is now known to be achieved by the release of hydrogen ions and proteases into the `resorption pit' beneath the ruffled border of the osteoclast \citep{hall,vaananen-zhao}. Interestingly,  observations on \textit{in vitro} osteoclasts suggest that they resorb bone matrix for a period of time, then detach from the bone surface, before continuing resorption at a different site of the bone surface \citep{hall,martin2002,vaananen-zhao}. However, it is known that \textit{in vitro} observations need to be interpreted with caution, as osteoclasts \textit{in vitro} often exhibit obviously different behaviours (such as low bone resorption activity), raising questions about their phenotype \citep{susa-etal}. Whether this detachment-delay-reattachment behaviour actually occurs \textit{in vivo} is not known \citep{everts-etal}, but this question is explored here using our computational model.

Serial histological sectioning of bone in humans and other animals has revealed that osteons are irregular cylindrical structures that vary in cross-sectional shape and anastomose extensively with one another, forming a complex network structure \citep{Cohen1958,tappen,stout-etal,moshin-etal}. More recently, microCT imaging has enabled the direct visualisation of the resorption cavity shape and of the network structure of Haversian canals in three spatial dimensions \citep{britz-etal,Cooper2006}. In particular, \cite{Cooper2006} found that the morphology of the \bmu\ spaces was varied, and included unidirectional, bi-direction and branched \bmu\ morphologies. More specifically, the cutting cones of the \bmu s imaged by \cite{Cooper2006} showed a generally ellipsoidal shape, but often the exact shape of the cutting cone is variable. In some image reconstructions shown in \citep{Cooper2006}, the surface of the cutting cone is noticeably rough. Our computational model enables us to hypothesise processes by which the network structure of osteons might emerge or that influence the size and shape of the resorption cavity.

After many decades of dedicated research by bone biologists, there now appears to be sufficient quantitative data and observations to develop a computational model describing the evolution of the cutting cone in a single \bmu\ in cortical bone. Our aim in this paper is to begin the process of integrating and assessing some of the reported data and observations identified above, and in doing so, start to develop a `dynamic picture' of osteoclast resorption and cutting cone evolution at the front of a moving cortical \bmu.

Computational models of bone remodelling by \bmu s have been proposed previously. For example, in \citep{buenzli-pivonka-smith,ryser-nigma-komarova}, the spatio-temporal organisation of bone cells within a single \bmu\ was investigated. Other works have focused on the external mechanical conditions of \bmu s \citep{Oers2008a,Oers2008b,Burger2003,garcia-aznar-etal}. Our model is similar to the discrete model of \cite{Oers2008a,Oers2008b}, which is based on a cellular Potts model \citep{glazier-graner}. In \citep{Oers2008b}, a relationship between osteon diameter and strain is investigated, but the generation and maintenance of osteoclasts is solely determined by the local strain energy density at the bone surface. Furthermore, that relationship is not explained in terms of properties of the constituents of a \bmu\ (such as number and activity of osteoclasts). Several biological processes influence osteoclastogenesis in a cortical \bmu. To our knowledge, our model is the first to address how important biological processes that support osteoclastic resorption in cortical bone (in particular, the growth of a blood vessel and osteoclastic nuclei renewal) influence the shape of a \bmu's cutting cone, osteon diameter and a \bmu's progression rate.

While the computational model developed here only involves osteoclasts and focuses on the formation of the resorption cavity at the front of a single cortical \bmu, it is helpful to be more specific about what the model does not include. The model does not include other cells in the \bmu\ (e.g. osteoblasts, osteocytes, endothelial cells or immune cells) and does not include other processes occurring in the \bmu\ (e.g. osteoid formation and mineralization, and osteocytic signals). Though much research in bone biology since the 1990s has been focused on discovering and understanding signaling pathways that coordinate cells and processes within a \bmu, the computational model developed here does not include explicit cell-cell signaling or explicit intracellular signal-transduction pathways. Specifically, the model does not include explicit signaling processes leading to osteoclast formation, or explicit signaling processes that may operate within the cutting cone. Rather, in this paper we have taken the approach of simulating key osteoclast behaviours that enable formation of a resorption cavity that are consistent with experimental observations. We simply establish a lattice-based computational model of osteoclast resorption, and identify some of the requirements that are particularly important for the development of normal resorption cavity shapes consistent with experimental observations.

The computational model is investigated through `parameter variations'. In much the same way that under-expression or over-expression of a gene illustrates the effect of that gene in an \textit{in vivo} system, so parameter variation illustrates the effect of that parameter in an \textit{in silico} system (that is, in the computational model). By this means, key influences on the outputs of the computational model can be identified.

The paper is organised as follows: in Section~\ref{model-description} we describe our lattice-based computational model for the evolution of the \bmu\ resorption cavity (a precise mathematical description of this model is given in Appendix~\ref{appx:model}). In Section~\ref{results} we first identify the importance of cell-cell adhesion, cell-bone adhesion and osteoclast access to the resorption surface for normal cavity shape formation, and then go on to identify the critical importance that the position of the tip of the blood vessel within the resorption cavity has on the evolution of the shape of the resorption cavity. The influence on resorption cavity shape and osteon morphology of osteoclast number, osteoclast longevity and nuclei fusion is also reported in Section~\ref{results}. In Section~\ref{discussion} we discuss these results and make predictions of osteoclast movement patterns within the resorption cavity, which to date have not been possible to observe experimentally. We also compare model predictions with experimental data on osteon shape and \bmu\ resorption cavity shape. Finally, in Section~\ref{conclusions}, we conclude and note future research directions.

\section{Model description\label{model-description}}
We first summarise current biological knowledge of osteoclastic resorption in cortical \bmu s before outlining the main features of the model.

Osteoclasts are known to resorb bone matrix to form the so-called `cutting cone' at the front of a cortical \bmu\ \citep{martin-burr-sharkey,parfitt1994}. To dissolve the bone matrix, an osteoclast first attaches to the bone surface, forming a `sealing zone' that encircles a small portion of the bone surface that will be resorbed \citep{hall,vaananen-zhao}. Within this sealed portion of the bone surface, hydrogen ions and proteases released through the osteoclast's `ruffled border' dissolves both the mineral and collageneous components of the bone matrix. \textit{In vitro} studies suggest that complete dissolution of the collageneous component of the bone matrix may require exposure to the extracellular microenvironment, where several proteases (such as \mmp s) secreted by other cells in the \bmu\ are known to be able to degrade this organic component \citep{everts-etal,vaananen-zhao}. While osteoclasts can also degrade this organic component through the production of cathepsins \citep{baron-etal}, they are observed \textit{in vitro} to physically detach from the bone surface before continuing their resorption at more highly mineralised bone locations \citep{hall,chambers-etal,matsuoka-etal,schilling-etal}. It is not known whether this behaviour occurs \textit{in vivo} \citep{martin2002,vaananen-zhao}. The resorption process is sustained by the growth of a blood vessel, which provides the local \bmu\ microenvironment with both nutrients and precursor cells (in particular precursor osteoclasts) \citep{parfitt1998}. The maturation of osteoclasts in a \bmu\ is the result of a cascade of events that take place around the tip of the blood vessel. This cascade of events involves several cell types and molecules (such as pre-osteoblasts, \mcsf, and \rankl, see e.g.\ \citep{roodman,martin}) and includes the fusion of mononucleated pre-osteoclasts. This leads to the generation of mature, multinucleated osteoclasts at the front of the \bmu\ \citep{parfitt1998}. Multinucleated osteoclasts, however, are believed to be dynamic entities. Autoradiographic studies of cell migration in cortical \bmu s have revealed that thymidine-labelled nuclei from osteoclast precursor cells fuse randomly with existing osteoclasts \citep{jaworski-duck-sekaly,miller,parfitt1994,bronckers-etal,fukushima-bekker-gay}. This process leads to the continual renewal of the nuclei within osteoclasts at the front of \bmu s, which is believed to account for an increase in the osteoclasts' lifespan and/or the persistence of their resorptive activity.

\begin{figure}
	\centering\includegraphics[width=\columnwidth]{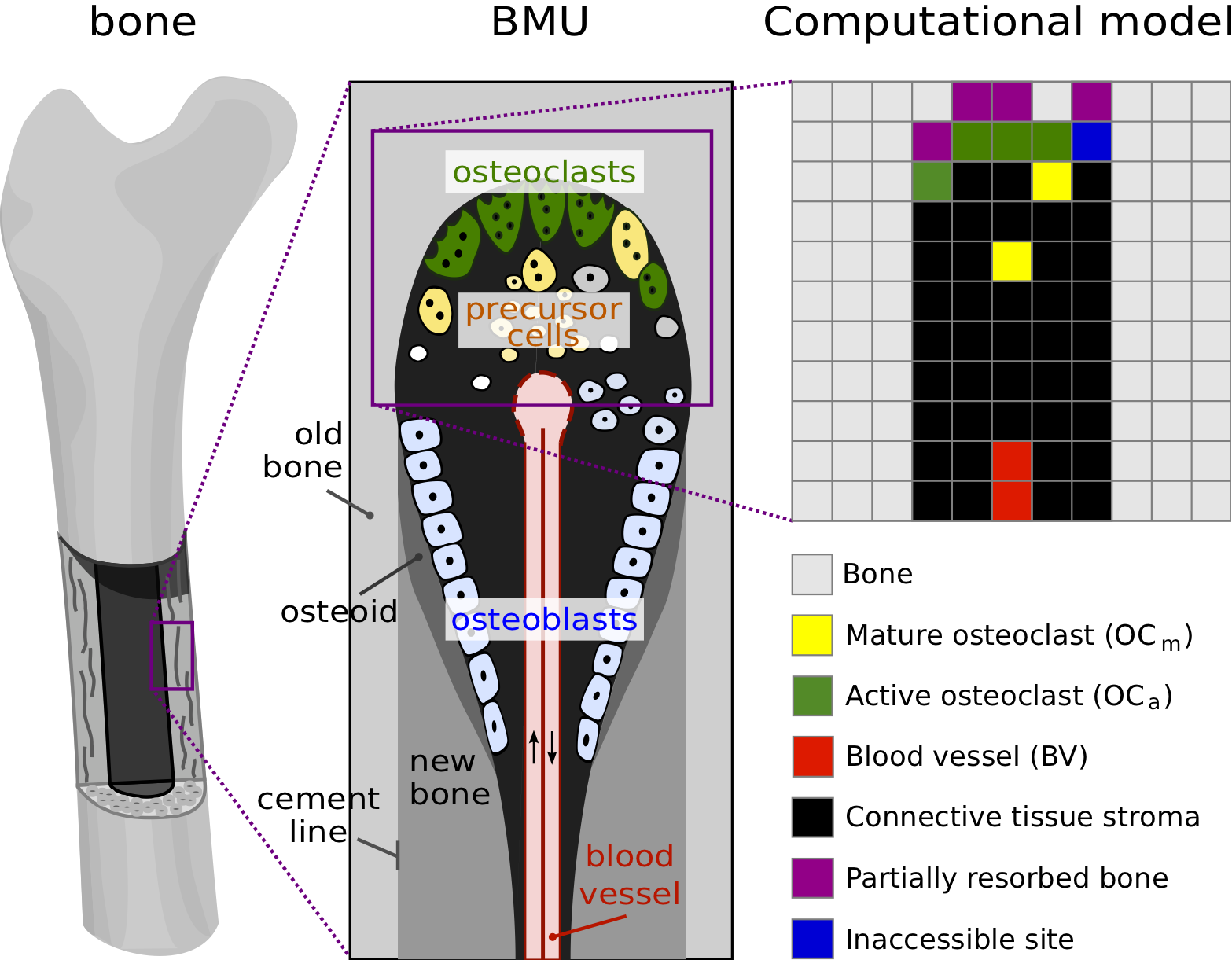}
	\caption{Schematic representation of a \bmu\ and two-dimensional lattice. Active osteoclasts are resorbing bone matrix at the front of a \bmu. They are generated through the fusion of mononuclear precursor cells provided by a blood vessel. In our two dimensional square lattice model, each lattice site denotes either bone (gray), partially resorbed bone (purple), resorbed but momentarily inaccessible bone (blue), an active osteoclast (green), a migrating mature osteoclast (yellow), blood vessel components (red), or loose connective tissue stroma (black).}
	\label{Fig:BMU}
\end{figure}
In this paper, we consider a two-dimensional slice of cortical bone being resorbed by osteoclasts, as would be seen in a thin longitudinal section running through the centreline of a cortical \bmu\ (Figure~\ref{Fig:BMU}). To investigate this system, we take a modelling approach in which each osteoclast is tracked individually and the position and activity of the osteoclasts (resorbing or not) is updated at regular time intervals. This type of modelling approach is known as `agent-based modelling', and has been widely used in recent years for computational models of tumours and other biological systems \citep{mansury-etal,walker-etal,zhang-deisboeck-etal,anderson-cummings-etal,gerlee-anderson,jeon-quaranta-cummings}. It differs from the more commonly used continuous approaches that use ordinary or partial differential equations, in that discrete cells are represented, rather than ``smeared'' quantities. The choice of this modelling approach is made here because: (i)~a \bmu\ contains only few osteoclasts confined within a very small volume; (ii)~our purpose is to simulate the individual and collective behaviours of osteoclasts, and to understand how these behaviours in turn influence bone resorption within a single \bmu. The consideration of individual osteoclasts in this agent-based approach allows us to represent complex biochemical mechanisms (such as osteoclast--bone adhesion) directly at the level of individual cells: the biochemical mechanisms are represented mathematically as `evolution rules' for the migration behaviour and resorption activity of the osteoclasts. To simplify the mathematical representation, we further assume that space is discretised: the individual osteoclasts in the \bmu\ are assumed to migrate on a two-dimensional lattice (see Figure~\ref{Fig:BMU}). The lattice step size $\sigma=40~\um$ is taken to correspond to the average size of multinucleated osteoclasts.

\begin{figure}
    \centering\includegraphics[width=\columnwidth]{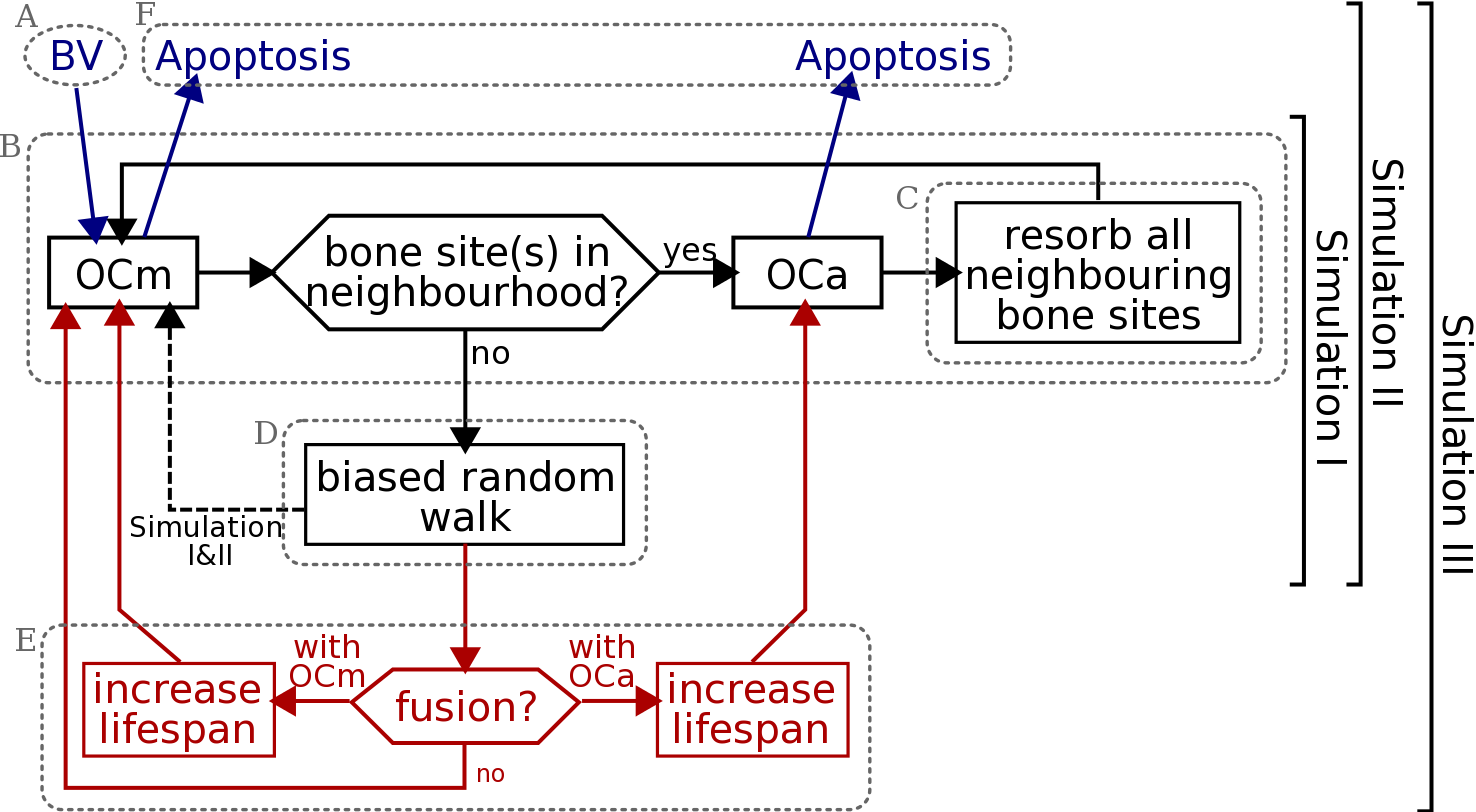}
    \caption{Flow chart of the model, where \ocm\ denotes a mature, migrating osteoclast, \oca\ denotes an actively resorbing osteoclast, and \bv\ stands for supply from a blood vessel. The encircled regions A--F correspond to the model features introduced in Section~\ref{model-description}. In Section~\ref{results}, Simulation I--III include incrementally more model features as shown.}
    \label{Fig:FlowChart}
\end{figure}

The biological processes of osteoclastic resorption summarised above are captured in our model by the following model features. A flow chart of these model features is shown in Figure~\ref{Fig:FlowChart}:
\begin{itemize}
    \item Fully mature osteoclasts are assumed to be generated at a rate $\eta_\oc$ (in $\da^{-1}$) from the tip of a blood vessel (\bv) which grows towards the front of the \bmu\ at a maximum rate $v_\bv$ (in $\um/\da$) (Figure~\ref{Fig:FlowChart}, A);
    \item An osteoclast can be in either of two states: `migrating' (not resorbing) or `active' (resorbing). We denote a migrating osteoclast by \ocm\ and an active osteoclast by \oca. An osteoclast becomes active as soon as it reaches a bone surface. It remains active and immotile until all of its surrounding bone sites are resorbed, at which point it becomes a migrating osteoclast again (Figure~\ref{Fig:FlowChart}, B);
    \item Dissolution of the bone matrix by an \oca\ is represented by a kinetic dissolution law that gradually reduces bone density in time. Bone matrix at a lattice site that has been resorbed by more than 90\% is assumed to become a cavity site filled with connective tissue stroma. However, such a site is deemed inaccessible to \ocm s for a time period $\tauclean$ to allow for possible extracellular collagen digestion (Figure~\ref{Fig:FlowChart}, C);
    \item The migration of \ocm s through the connective tissue stroma is modelled as a `biased random walk'. At each time increment, the \ocm\ chooses a lattice site to migrate to with a probability that depends on the presence of other components in its vicinity. The interaction of the \ocm\ with these components is described by `interaction energies'. For example, $E_{\oc-\oc}$ denotes an osteoclast--osteoclast adhesion energy, and $E_{\oc-\bone}$ denotes an osteoclast--bone adhesion energy. The lower the total energy at a neighbouring lattice site, the higher the probability of migrating towards this site (Figure~\ref{Fig:FlowChart}, D);
    \item Osteoclasts that are newly generated are initially assigned a fixed lifespan $\tau_\oc$. To account for a lifespan-increasing nuclei renewal process in our simulations, a migrating \ocm\ can fuse with an existing \oca\ or \ocm\ (with different probabilities, depending on the `fusion energies' $E_{\ocm-\oca}^\text{fusion}$ and $E_{\ocm-\ocm}^\text{fusion}$, respectively). The lifespan of the osteoclast resulting from this fusion is increased by the remaining lifetime of the fusing \ocm\ (Figure~\ref{Fig:FlowChart}, E);
    \item When the age of an osteoclast reaches its alloted lifespan (whether that lifespan has been increased by nuclei renewal or not), the cell is removed from the system (Figure~\ref{Fig:FlowChart}, F).
\end{itemize}
The technical details of the model are discussed in Appendix~\ref{appx:model}, and a full list of the model parameters is given in Table~\ref{Table:Parameter}. For each parameter, Table~\ref{Table:Parameter} also lists the range of values investigated in Section~\ref{results}, and a so-called `default' value, which is assigned unless that parameter is explicitly varied.

\section{Simulation results\label{results}}
A multiplicity of physiological events takes place in the resorption cone of a \bmu\ and it can be difficult to gain an understanding of the specific influence of each of these events on the dynamics of bone resorption in the \bmu. However, some of these influences can be studied using our model. To this end, we develop an understanding of the model by including model features incrementally. We denote this incremental introduction of model features by \m1, \m2 and \m3, in increasing order of complexity. The model features taken into account in each of these simulations are highlighted in Figure~\ref{Fig:FlowChart}.\footnote{It is noted that \m1 and \m2 can be retrieved from the full model (\m3) by considering extreme values for some parameters. Specifically, \m2 is obtained from \m3 by impeding nuclei renewal by fusion ($E_{\ocm-\ocm}^\text{fusion} = E_{\ocm-\oca}^\text{fusion}=\infty$) and \m1 is obtained from \m2 by assuming non-renewed osteoclasts ($\eta_\oc=0, v_\oc=0, \tau_\oc=\infty$).}

A key model output that is used to narrow down a physiologically realistic parameter space is the shape and extent of the resorption cavity after a simulation of 30 days. This choice of the duration of the simulation enables the system to evolve from its initial conditions and to reach a pseudo steady state at the front of the \bmu, where the shape of the cutting cone as well as the relative spatial distribution of cells is no longer changing over time on average. This simulation duration is also well within the average lifespan of a \bmu, which is estimated to be about 6--12 months \citep{parfitt1994}.

\subsection{\m1}
\m1 has the fewest model features (Figure~\ref{Fig:FlowChart}). In \m1, osteoclasts are neither dynamically generated nor eliminated. The total number of osteoclasts is fixed and their lifespan is assumed to be infinite (no blood vessel nor nuclei renewal mechanism is considered for this study). While this simulation is clearly not physiological, it allows us to develop an understanding of how osteoclast--osteoclast interaction, osteoclast--bone adhesion and extracellular collagen digestion (via the parameters $E_{\oc-\oc}$, $E_{\oc-\bone}$ and $\tauclean$, respectively) influence the dynamics of bone resorption, without the additional complexity of birth and death processes of the osteoclasts. The initial configuration for the simulation consists of nine osteoclasts placed at the nine central sites of the cavity shown in Figure~\ref{Fig:Neighbor}(a).

Interestingly, a unidirectional resorption pattern (resulting in an elongated cavity) spontaneously emerges for an osteoclast--osteoclast adhesion much stronger than an osteoclast--bone adhesion ($E_{\oc-\oc} \ll E_{\oc-\bone} < 0$). However, one expects the physiological situation to be the reverse case, where osteoclast--bone adhesion is strong compared to an osteoclast--osteoclast interaction ($E_{\oc-\bone} \ll E_{\oc-\oc} \leq 0$). Indeed, it is well established that mature osteoclasts are found in the vicinity of bone surfaces and can strongly adhere to bone through the formation of a dense actin ring (the sealing zone) \citep{vaananen-zhao}. But mature osteoclasts do not seem to strongly adhere to one another through the expression of specific adhesion molecules after their formation, though there are possibly some cadherin molecules on the mature osteoclast surface \citep{helfrich-etal,civitelli-etal}. To represent this fact, and because mature osteoclasts are not usually found isolated in cortical \bmu s, a small osteoclast--osteoclast interaction energy $E_{\oc-\oc}$ is selected (see Table~\ref{Table:Parameter}).  In our simulations, a relatively strong osteoclast--bone adhesion energy $E_{\oc-\bone}$ makes bone resorption more persistent. At weaker values of $E_{\oc-\bone}$, the osteoclast spends most of its time migrating within the connective tissue stroma and may be found far from a bone surface.

\begin{figure}
    \hspace{2mm}\includegraphics[width=0.92\columnwidth]{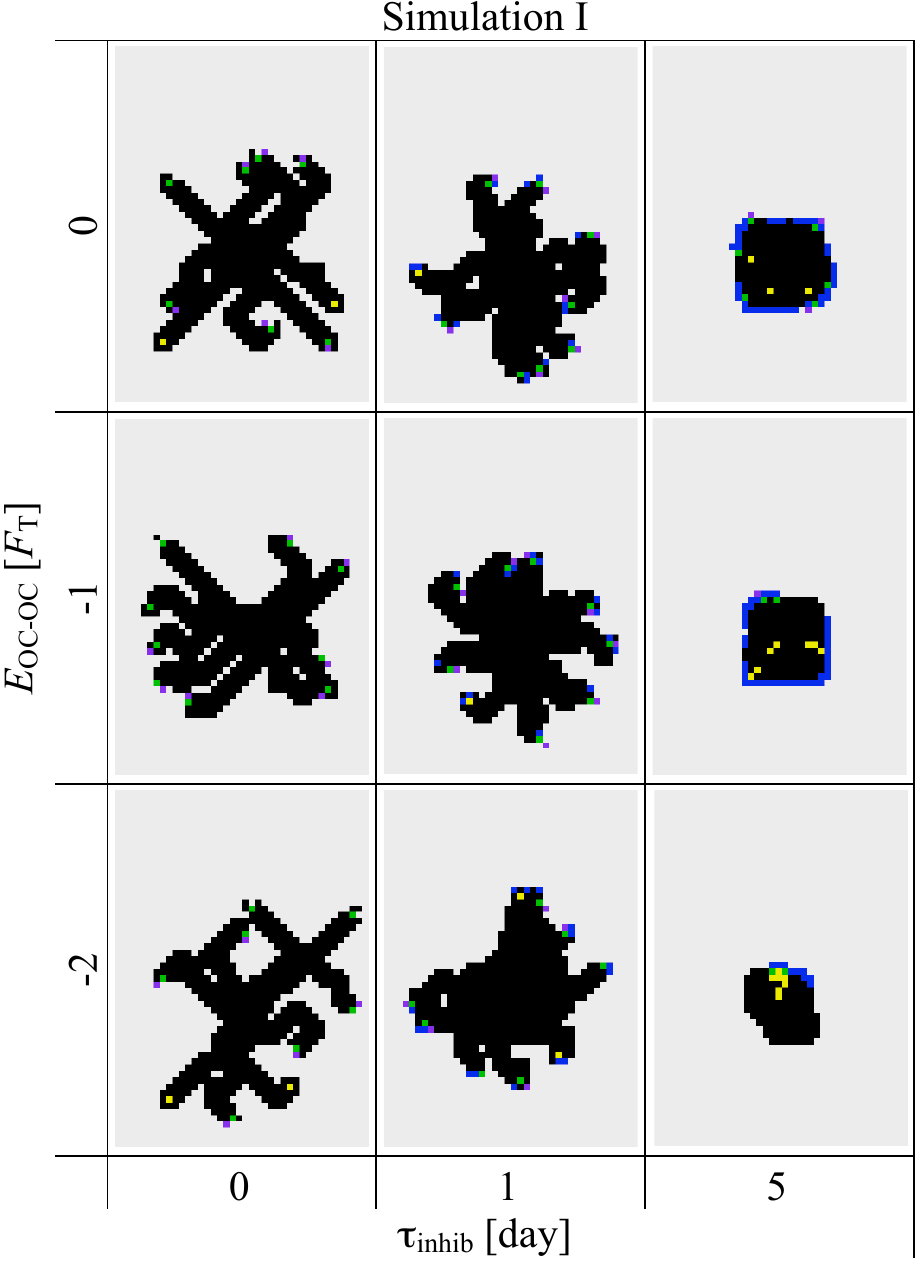}
    \caption{\m1. Snapshots of the lattice at $t=30~\days$ showing the effect of osteoclast--osteoclast interaction ($E_{\oc-\oc}$) and extracellular collagen digestion period ($\tauclean$). The portion of the lattice shown in these snapshots is a rectangle of $40\times56$ sites, corresponding to $1600~\um\times 2240~\um$. (Colour code as in Figure~\ref{Fig:BMU}.)}
	\label{Fig:Model1}
\end{figure}
Figure~\ref{Fig:Model1} shows the combined effect of osteoclast--osteoclast interaction and extracellular digestion time on the resorption cavity. Preventing osteoclast attachment for a period $\tauclean$ for extracellular collagen digestion strongly affects the behaviour of bone resorption. With increasing values of $\tauclean$, the osteoclast migrates for longer periods of time. At smaller values of $\tauclean$, several osteoclasts are observed to resorb bone matrix side by side at the beginning. But they move apart from each other and independently resorb bone matrix at later times by branching out from the original resorption cavity and go their separate ways. Increasing the strength of osteoclast-osteoclast interaction $E_{\oc-\oc}$ slightly reduces the occurrence of branching. Collagen digestion time $\tauclean$ strongly affects branching behaviour, but this parameter also strongly influences the efficacy of bone resorption, as can be seen in Figure~\ref{Fig:Model1} for $\tauclean=5~\days$.

From Figure~\ref{Fig:Model1}, one also sees that bone resorption in our model occurs preferentially in diagonal directions. Due to the relatively strong osteoclast--bone adhesion, \ocm s preferentially maximise contact with bone surfaces, leading to diagonal movement on the lattice, since bone surface at a corner is larger than bone surface along a flat edge. Such a diagonal bias introduces an unavoidable anisotropy from a lattice model. However, this shows that the local curvature of the bone surface directly affects osteoclastic bone resorption.

\begin{figure*}
    \hspace{-3.5mm}
    \centering\begin{tabular}{ll}
    {\small (a)}&\hspace{8mm}{\small (b)}
    \\
    \includegraphics[width=0.92\columnwidth]{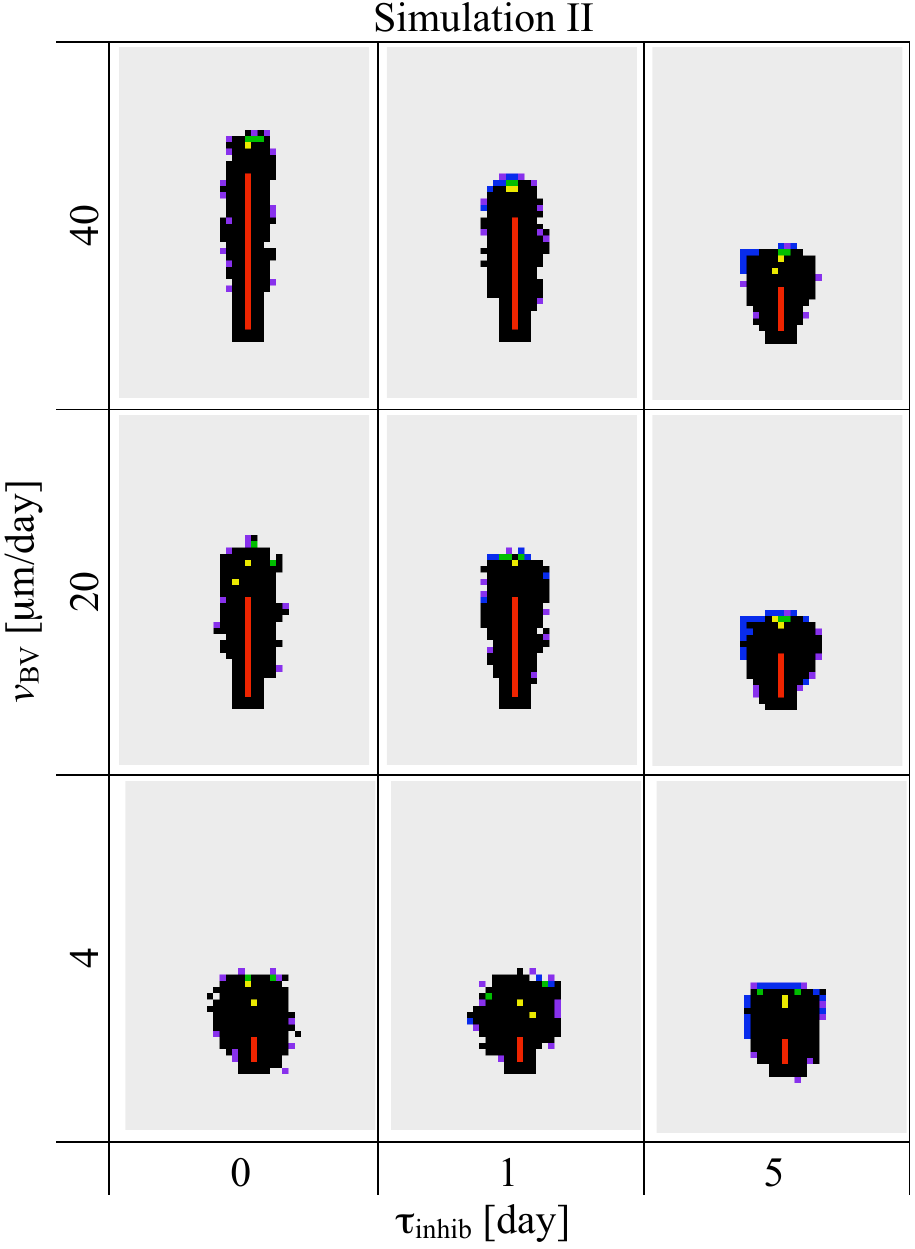} &
    \hspace{8mm}\includegraphics[width=0.92\columnwidth]{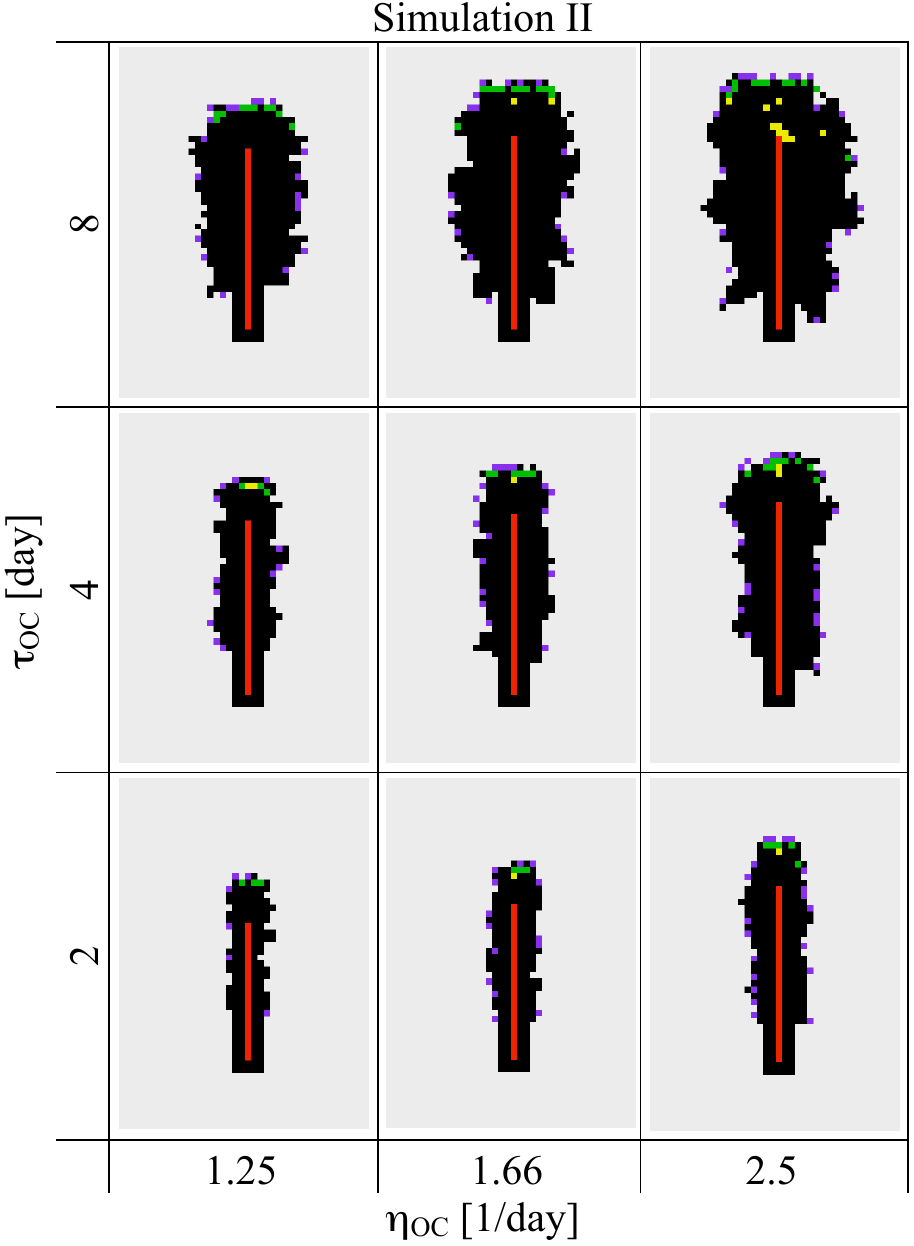}
    \end{tabular}
    \caption{\m2. Snapshots of a $1600\,\um\times 2240\,\um$ portion of the lattice at $t=30\, \days$. (a) Effect of maximum rate of growth of the blood vessel ($v_\bv$) and extracellular collagen digestion period ($\tauclean$); (b) Effect of generation rate of osteoclasts ($\eta_\oc$) and lifespan of osteoclasts ($\tau_\oc$). (Colour code as in Figure~\ref{Fig:BMU}.)}
    \label{Fig:Model2}
\end{figure*}

Our results from \m1 suggest that the magnitude of osteoclast--bone adhesion energy $E_{\oc-\bone}$ should be chosen large compared to $E_{\oc-\oc}$ so that bone resorption is more persistent and to avoid osteoclasts being far from a bone surface. As discussed above, these parameter values are in fact expected physiologically, but importantly, we observe that by themselves, they do not lead to a unidirectional progression of the resorption cavity. Rather, they lead to directionless openings that grow larger over time. It is apparent that the parameter $\tauclean$ (extracellular collagen digestion period) needs to be fairly small, as a large value strongly reduces the amount of bone resorption possible (see Figure~\ref{Fig:Model1}; $\tauclean=5~\days$).

\subsection{\m2}
The absence of a preferred direction for the resorption of bone in \m1 is a clear indication that the model requires additional processes to produce a `cutting cone'. Such an outcome can be obtained by including the growth of a blood vessel. This model feature can account for both the renewal of the osteoclast population and the unidirectional progression of the \bmu. Three new parameters are introduced in \m2, namely: the maximal rate of growth of the blood vessel ($v_\bv$), the generation rate of osteoclasts ($\eta_\oc$) and the lifespan of osteoclasts ($\tau_\oc$). The initial configuration of the simulation is depicted in the appendix (Figure~\ref{Fig:Neighbor}(b)). Experimentally, osteons have a diameter of the order of 200--350 \um\ and \bmu s progress through bone at a rate of 20--40 \um/\da\ \citep{parfitt1983b,parfitt1994,robling-castillo-turner}. To retrieve such osteon diameters and \bmu\ progression rates in our simulations, a combination of both values of $v_\bv$ of 20--40 \um/\da\ and small values of $\tauclean$ needs to be chosen (see Figure~\ref{Fig:Model2}(a)). For such small values of $\tauclean$, osteoclasts are not observed to migrate away from a previous resorption site before starting to resorb again. This suggests that `detachment--delay--reattachment' behaviour of an osteoclast is very localised within a cortical \bmu\ (with no significant migration before reattachment). We infer that extracellular collagen dissolution is unlikely to be a rate limiting step for resorption within a cortical \bmu.
\begin{figure}[t]
    \centering{\includegraphics[width=0.7\columnwidth]{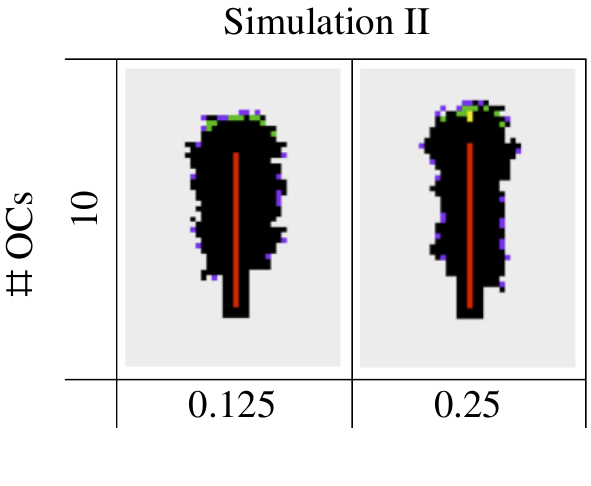}\\
    \vspace{-6mm}\includegraphics[width=0.7\columnwidth]{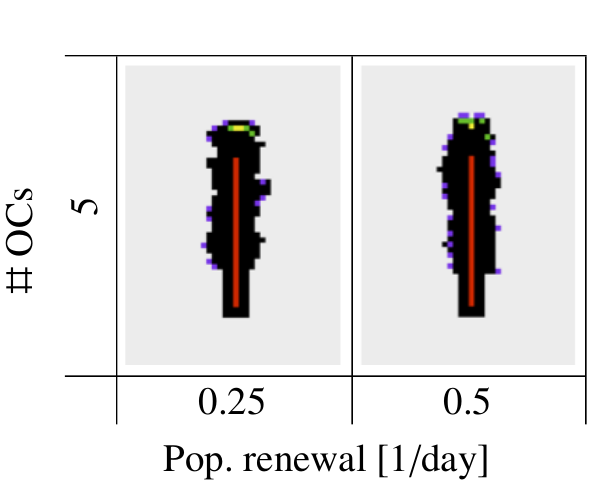}}
    \caption{\m2. Four snapshots of the lattice from Figure~\ref{Fig:Model2} are reproduced here in dependence upon average number of osteoclasts and osteoclast population renewal rate (see text).}
    \label{Fig:Model2-nbOCs-vs-popRenewalRate}
\end{figure}
 
Figure~\ref{Fig:Model2}(b) shows the combined effect of the generation rate of osteoclasts ($\eta_\oc$) and of the lifespan of osteoclasts ($\tau_\oc$) on the resorption cavity. These model parameters affect both the average number of osteoclasts in the cavity (which is equal to $\eta_\oc \tau_\oc$) and the renewal rate of this population (which is equal to $1/\tau_\oc$). The generation rate and lifespan of osteoclasts are often disrupted in bone diseases \citep{manolagas}, and they may also slightly vary locally or in different bone sites, leading to local inhomogeneities of osteon morphology or \bmu\ cutting cone shape (see Section~\ref{discussion}). Not surprisingly, osteon diameter strongly depends on the number of osteoclasts in the cavity: it increases as either $\eta_\oc$ or $\tau_\oc$ (or both) increases in Figure~\ref{Fig:Model2}(b). The renewal rate of the osteoclast population for a given number of osteoclasts only marginally affects osteon diameter. However, this renewal rate modifies the roughness of the cavity surface. In Figure~\ref{Fig:Model2-nbOCs-vs-popRenewalRate}, we reproduce four snapshots of the lattice from Figure~\ref{Fig:Model2}, but organise them according to the average number of osteoclasts and osteoclast population renewal rate. The snapshots reproduced correspond to the snapshots of Figure~\ref{Fig:Model2} obtained for $(\eta_\oc,\tau_\oc)=(1.25/\da, 4~\days)$, and $(2.5/\da, 2~\days)$, which both have an average number of $5$ osteoclasts but population renewal rates of $0.25/\da$ and $0.5/\da$, respectively, and the snapshots obtained for $(\eta_\oc,\tau_\oc)=(1.25/\da,8~\days)$, and $(2.5/\da,4~\days)$, which both have an average number of $10$ osteoclasts and population renewal rates of $0.125/\da$ and $0.25/\da$, respectively. A smaller population renewal rate is seen to induce slightly more ragged cavities. Indeed, low renewal rates allow the long-living osteoclasts to resorb for a long time away from the front of the \bmu. This leads to budding of resorption cavities (and thus rougher surfaces), particularly as seen in the snapshots of the lattice with $\tau_\oc=8~\days$ in Figure~\ref{Fig:Model2}.

Our results from \m2 are consistent with the expectation that the number of osteoclasts in a \bmu\ significantly affects the osteon diameter \citep{martin-pickett-zinaich,Oers2008b,britz-etal}. Our simulations suggest in addition that osteoclast population renewal rate influences the roughness at the osteonal cement line. It would be interesting to compare this prediction of our model with experimental data, but to our knowledge, there are no experimental studies investigating how roughness at the cement line of osteons may vary in various bone diseases or drug treatments targetting osteoclasts.

\subsection{\m3}
The life history of osteoclasts (from their generation to their apoptotic death) is a fundamental feature influencing bone resorption \citep{parfitt-etal1996,manolagas}. The lifespan and/or activity of osteoclasts is believed to be dynamically influenced by the renewal of their nuclei. However, it is not yet experimentally possible to follow the paths of single osteoclasts \textit{in vivo}, and so where and how osteoclasts are degraded remains unclear \citep{parfitt-etal1996,bronckers-etal,boyce-etal2002,Burger2003}. To investigate the interplay between osteoclast removal and nuclei renewal, we include in \m3 the possibility that \oca s renew their nuclei (and thereby increase their lifetime) through fusion with \ocm s \citep{fukushima-bekker-gay}. In \m3, all the model features presented in Section~\ref{model-description} now play a significant role, except for extracellular collagen digestion period $\tauclean$, which is set to $0~\days$, and for fusion between two \ocm s, which is ruled out (see~Appendix~\ref{appx:model}).

\begin{figure}
	\centering\includegraphics[width=.85\columnwidth,clip=true,hiresbb=true]{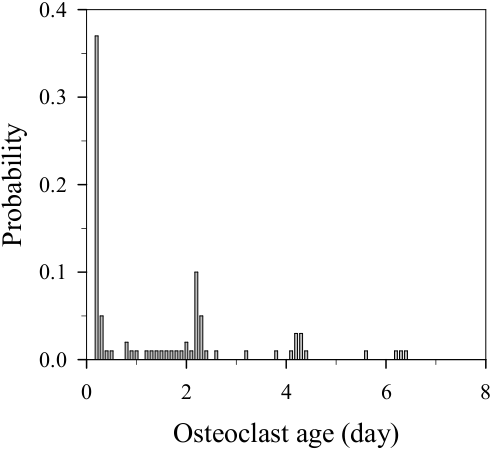}
    \caption{\m3. Probability distribution of osteoclast age at the time of their apoptotic death or their fusion into another osteoclast.}
    \label{Fig:CellAgeDistribution}
\end{figure}
Figure~\ref{Fig:CellAgeDistribution} shows the probability distribution of the age reached by osteoclasts in our model before their removal from the \bmu, whether by apoptotic death or by fusion into an existing \oca. This probability distribution thus represents the likelihood for a nuclei renewal process to occur in our simulations. In \m2, every osteoclast undergoes apoptosis at an age exactly equal to $\tau_\oc=2~\days$ and such a probability distribution is uniquely peaked at $2~\days$. By contrast, in \m3 the various peaks reflect modification of the lifespans upon nuclei renewal. Any peak before 2~days corresponds to \ocm s that are removed by fusion with an existing \oca. Any peak after 2~days corresponds to \oca s whose nuclei have been renewed once or several times. It is seen that the most probable fate of \ocm s is to fuse with an existing \oca\ shortly after their generation, at $0.2\, \days$ of age.  Due to the observed decay in peak height with age, the probability for an \oca\ of undergoing a nuclei renewal process decreases with time in our model. This is because long-lived osteoclasts gradually move further away from the location where new \ocm s are generated, making fusion less likely (see Section~\ref{discussion} for further discussion).

\begin{figure}
	\centering\includegraphics[width=\columnwidth]{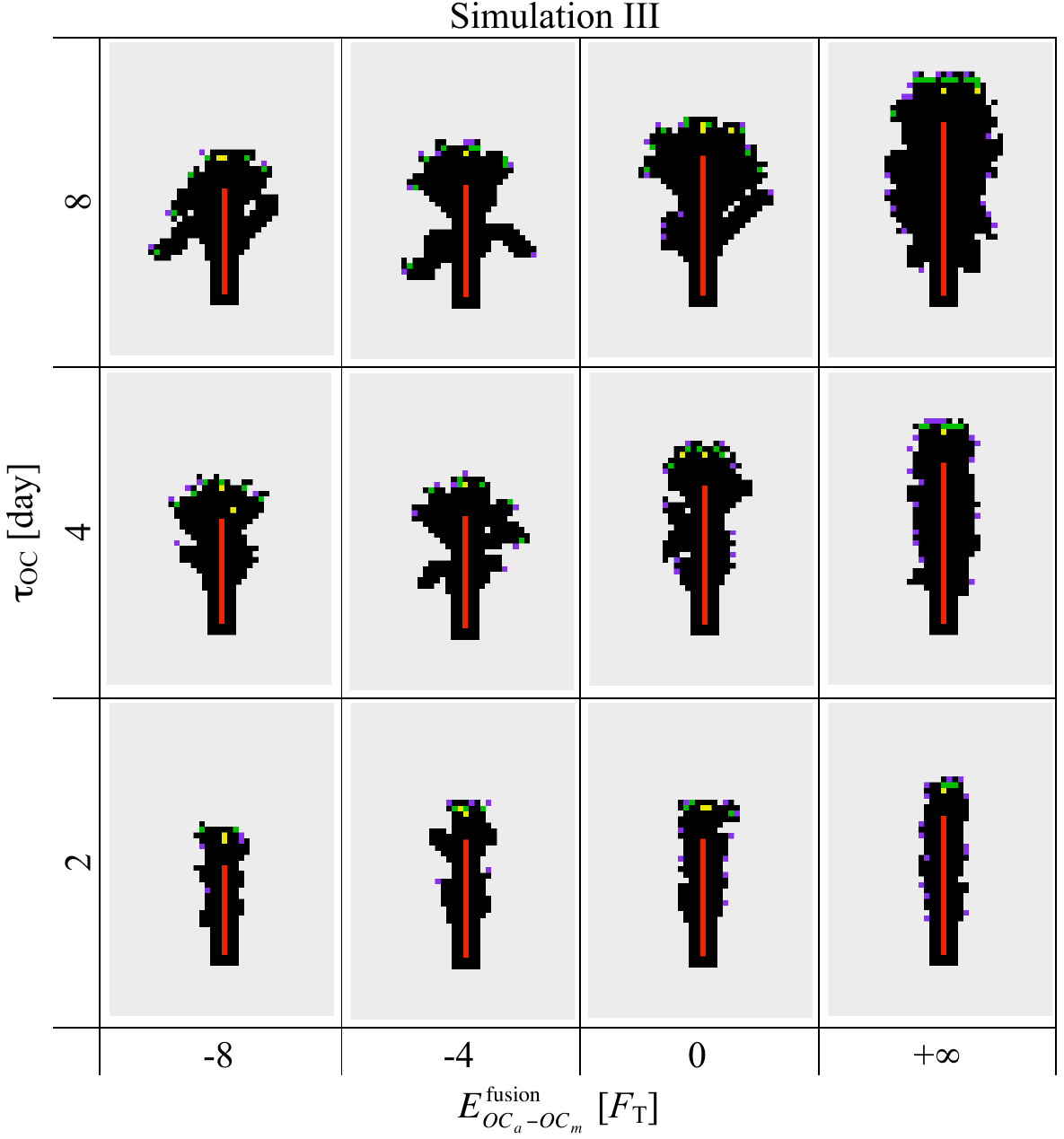}
	\caption{\m3. Snapshots of a $1600\,\um\times 2240\,\um$ portion of the lattice at $t=30\, \days$. When $E_{\ocm-\oca}^\text{fusion}=+\infty$, no fusion process takes place and the snapshots from \m2 (second column of Figure~\ref{Fig:Model2}(b)) are reproduced. (Colour code as in Figure~\ref{Fig:BMU}.)}
	\label{Fig:Model3}
\end{figure}
\begin{figure}
    \hspace{-4.5mm}
    \begin{tabular}{ll}%
    {\small \hspace{8mm}(a) Osteon diameter}& \hspace{-4mm}{\small (b) Cavity roughness}%
    \\%
    \includegraphics[width=0.56\columnwidth]{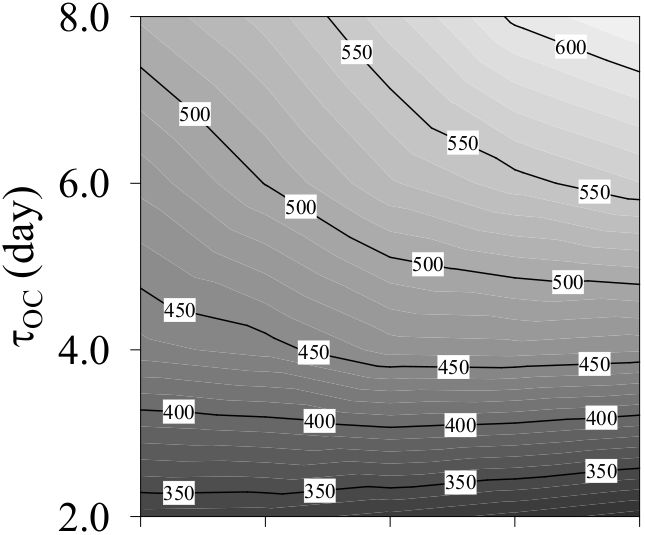}&%
    \hspace{-3mm}\includegraphics[width=0.45\columnwidth]{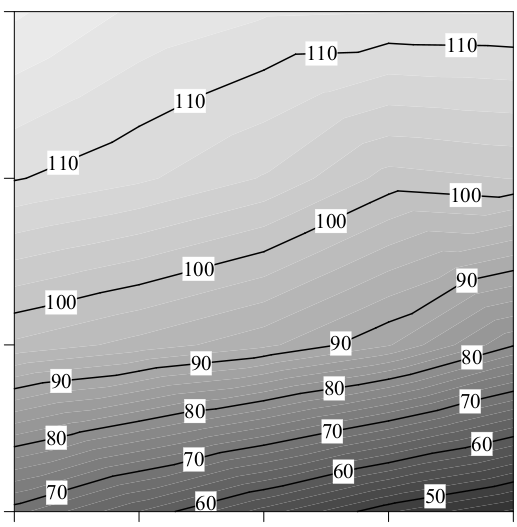}%
    \\%
    {\small \hspace{8mm}(c) Mean resorption rate/\oc}& \hspace{-4mm}{\small (d) \bmu\ progression rate}%
    \\%
    \includegraphics[width=0.565\columnwidth]{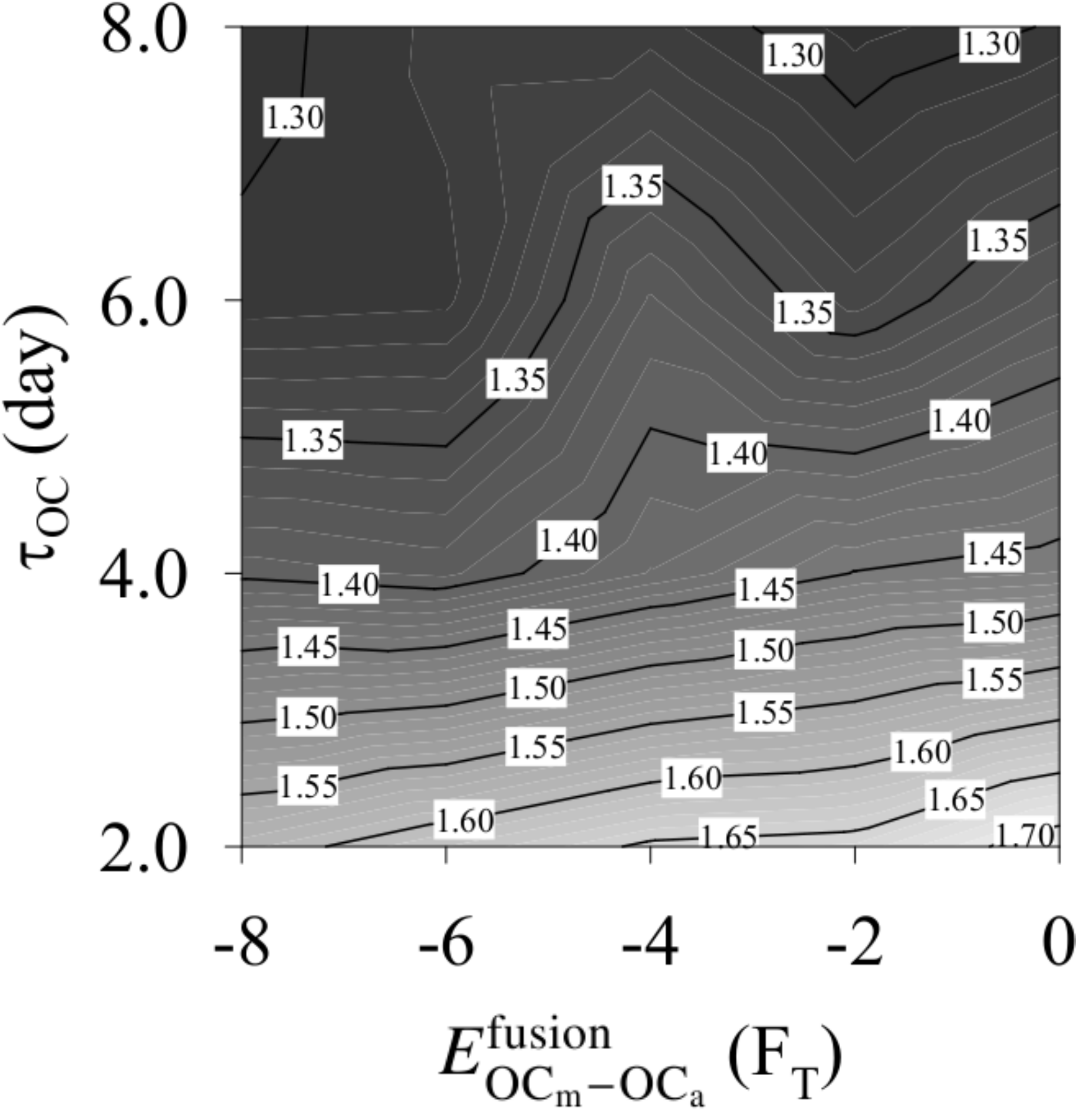}&%
    \hspace{-4mm}\includegraphics[width=0.46\columnwidth]{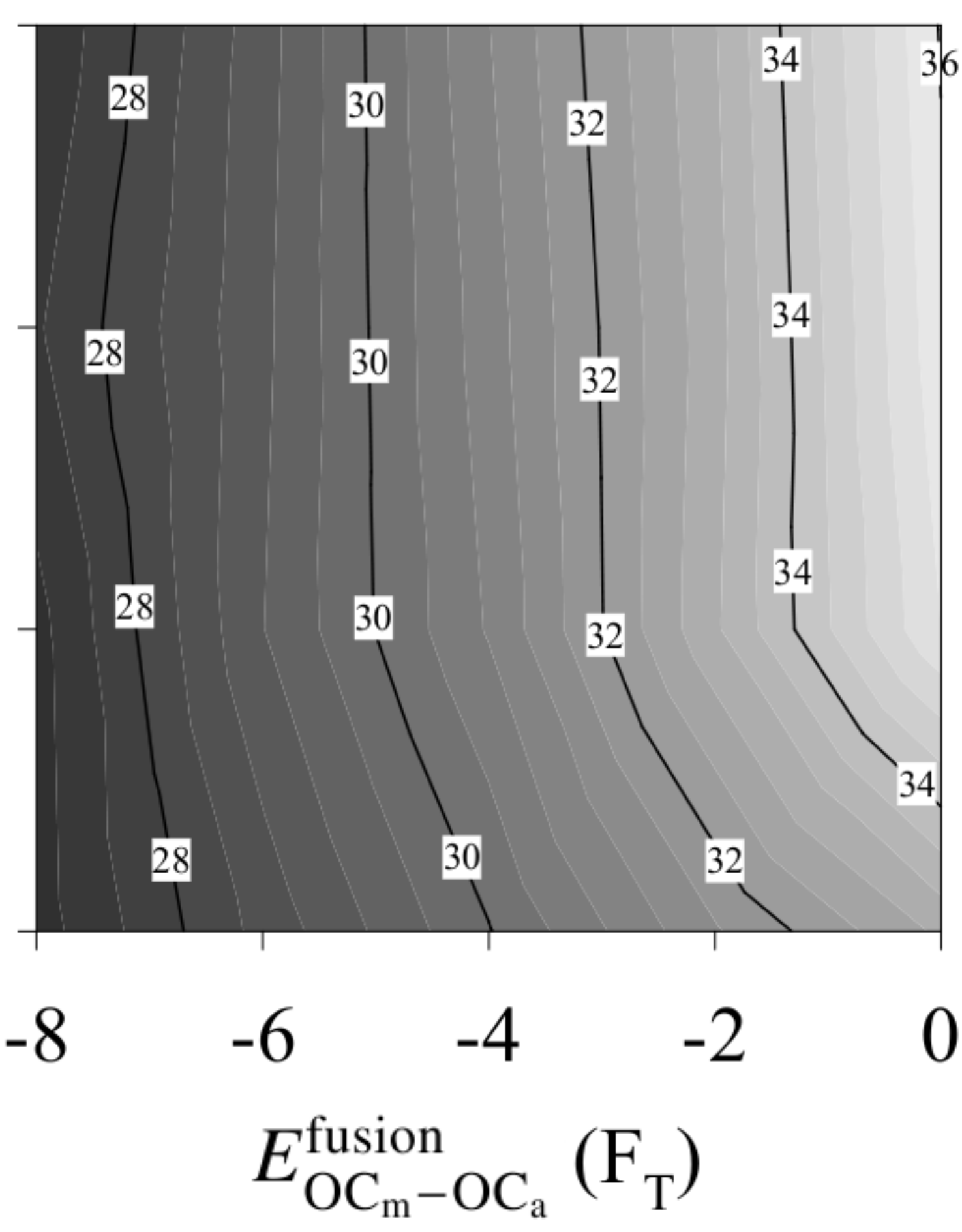}%
    \end{tabular}
    \caption{\m3. Some average resorption properties of the model are shown in dependence upon fusion affinity ($E_{\ocm-\oca}^\text{fusion}$) and lifespan of osteoclasts ($\tau_\oc$). (a) Osteon diameter (cavity width, averaged over rows of the lattice) [$\um$]; (b) Surface roughness (standard deviation of osteon diameter) [$\um$]; (c) Mean resorption rate per osteoclast (number of resorbed bone sites divided by the number of created osteoclasts and by the duration of the simulation) [$\da^{-1}$]; (d) \bmu\ progression rate (cavity length averaged over columns of the lattice divided by duration of the simulation) [$\um/\da$].}
    \label{Fig:Fusion-JF-vs-AT-data}
\end{figure}

Figure~\ref{Fig:Model3} shows that there is a striking difference in osteon morphology and osteon diameter between increasing the lifespan of osteoclasts through $\tau_\oc$ (the initial lifespan of \ocm s at their birth) or through nuclei renewal. Indeed, while increasing $\tau_\oc$ increases the average osteon diameter as in \m2 (see columns of Figure~\ref{Fig:Model3}), the strength of fusion affinity ($E_{\ocm-\oca}^\text{fusion}$) impacts the morphology of the osteon more strongly than its diameter (see rows of Figure~\ref{Fig:Model3}). This can also be appreciated in Figure~\ref{Fig:Fusion-JF-vs-AT-data}, where the influence of $E_{\ocm-\oca}^\text{fusion}$ and $\tau_\oc$ on the average osteon diameter, on the cavity surface roughness, on the \bmu\ progression rate, and on the mean resorption rate per osteoclast is shown.\footnote{For the calculation of these quantities, the initial rectangular cavity, still visible in the lower part of the snapshots in Figure~\ref{Fig:Model3}, is discarded.}

From the snapshots of the lattice in Figure~\ref{Fig:Model3} at constant $\tau_\oc$, nuclei renewal appears to support the emergence of `budding branches' from the main resorption cavity. The osteonal structures in humans and other animals is known to branch and anastomose into a network structure \citep{Cohen1958,tappen,stout-etal,moshin-etal,Cooper2006}. In our simulations, these branches do not progress further as we do not model branching of the central blood vessel, so osteoclasts are not renewed in the buds.

An interesting observation from our simulation results is that the \bmu\ progression rate is a poor indicator of the mean resorption rate per osteoclast (and vice-versa). Comparing Figures~\ref{Fig:Fusion-JF-vs-AT-data}(c) and (d), one sees that the \bmu\ progression rate is mainly independent of $\tau_\oc$ (the initial lifespan of \ocm s at their birth) but depends on $E_{\ocm-\oca}^\text{fusion}$ (the fusion energy), while the reverse is true of the mean resorption rate per osteoclast. These results show that the nuclei renewal process plays an important role in how fast a \bmu\ may resorb a path through bone, indicating that the trajectories of the individual osteoclasts within the resorption cone are influenced by the nuclei renewal process.

Our results from \m3 suggest that an increase in lifespan of osteoclasts by nuclei renewal could possibly lead to different typical trajectories of the osteoclasts in the cutting cone of the \bmu, and so may influence the occurrence of branching. This nuclei renewal process has, to our knowledge, not been considered in a computational model previously. The systematic study of cavity properties as well as the age probability distribution of the osteoclasts may provide ways to deduce how often nuclei renewal in an osteoclast occurs before it undergoes apoptosis.

\section{Discussion\label{discussion}}
Osteoclast resorption behaviour in cortical \bmu s is complex. Histological serial sectioning and microCT technology have revealed branching osteon structures, and have identified the non-uniformity between individual osteons and \bmu s \citep{Cohen1958,tappen,stout-etal,moshin-etal,Cooper2006}. Osteons and \bmu\ cutting cones may have different diameters, morphologies, and roughnesses. While these structural features of osteons and \bmu s are revealed by these experimental techniques, no information on the underlying cellular processes coordinating osteoclast movement and resorption behaviour is obtained. But it is the behaviour of osteoclasts and their interaction with each other and the bone matrix that create these complex network structures. In this paper, we have developed a novel computational model to gain a deeper understanding about the way osteoclast properties and cell-cell and cell-bone interactions can lead to different structural features of resorption cavities.

\begin{figure}
    \centering\includegraphics[width=\columnwidth]{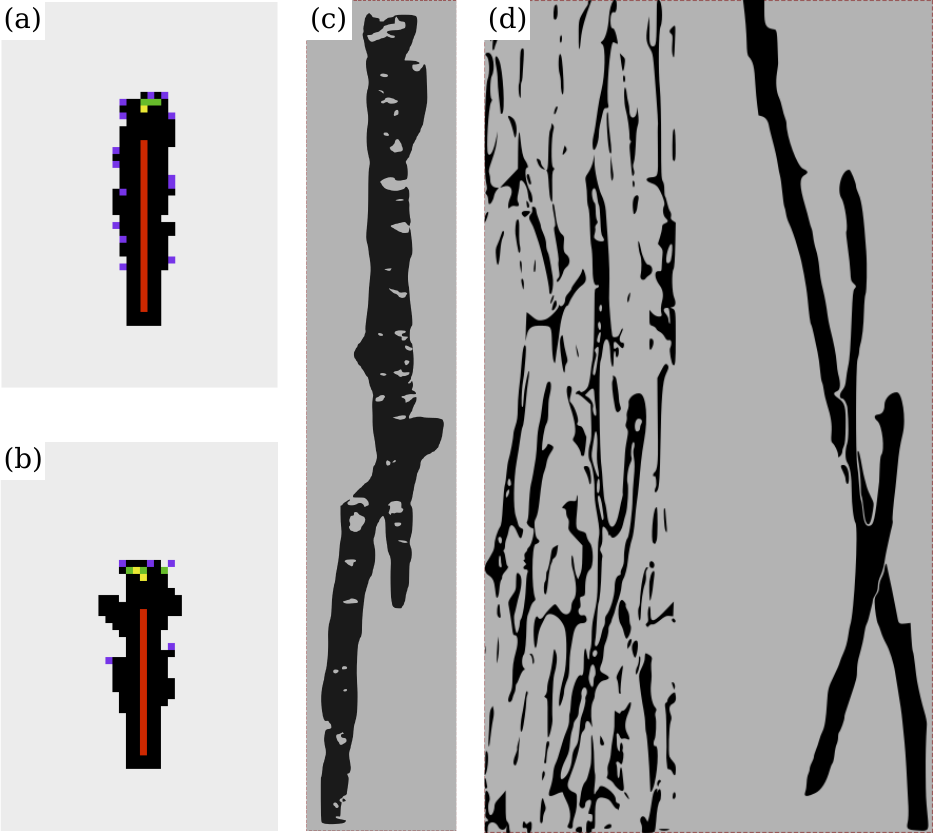}
    \caption{Comparison of simulation results with experimental images. (a) \m2, with default parameter values; (b) \m3, with the same parameter values but taking into account osteoclastic nuclei renewal; (c) Osteonal morphology for dogs, which closely resembles human osteons, redrawn from \citep{Cohen1958}; (d) BMU cutting cone morphology and Haversian canal network for humans, redrawn from \citep{Cooper2006} (see original for colours).}
    \label{fig:exp-images}
\end{figure}
We found that osteoclast--bone adhesion has to be strong to prevent osteoclasts from migrating far from the bone surface and that extracellular collagen digestion time ($\tau_\oc$) is unlikely to be a rate limiting step in osteoclastic bone resorption. In our model, a critical parameter determining unidirectionality of osteons is the growth direction of the blood vessel.

In Figure~\ref{fig:exp-images}, we compare osteonal shapes obtained from \m2 and \m3 with experimental data from \cite{Cohen1958} and \cite{Cooper2006}. Qualitatively our simulation results resemble the experimentally observed osteonal structures. For a suitable choice of model parameters osteonal diameters are in the range of experimentally observed diameters, i.e. $\approx200$--$350~\um$ \citep{parfitt1983b,robling-castillo-turner}. The osteon diameter is found to be strongly dependent on the average number of osteoclasts in the resorption cavity. Furthermore, we found that the osteoclast population renewal rate is related to the roughness of the osteon boundary at the cement line. Looking carefully at Figure~\ref{fig:exp-images}(c) one can see that the cement line is ragged, reflecting nonuniformity in the osteoclast resorption process, as in Figures~\ref{fig:exp-images}(a) and (b) for our model. Figure~\ref{fig:exp-images}(d) clearly shows the complex network stucture of Haversian systems (grey) and a branching secondary osteon (red). The osteoclast nuclei renewal process suggested in our model seems to support the occurrence of branching, as can be seen by comparing Figures~\ref{fig:exp-images}(a) and (b). Figures~\ref{fig:exp-images}(a) and (b) are obtained with the same parameter values except that nuclei renewal is taken into account in Figure~\ref{fig:exp-images}(b). We do not account for branching of the blood vessel, so buds of the cavity do not evolve into progressing branches. Still, osteoclast nuclei renewal is seen to have a distinctive effect on osteonal morphology, which is partly due to osteoclasts living longer but also due to osteoclasts taking different trajectories within the \bmu\ cavity. We now consider osteoclast trajectories within a \bmu.

Nuclei renewal in \m3 occurs through a fusion process, but fusion only takes place if two osteoclasts are found in the same region of space. This mechanism for lifespan extension is thus conditional and local. An osteoclast's nuclei are more likely to be renewed in the vicinity of the tip of the blood vessel, where new \ocm s are generated. But an osteoclast's nuclei may also be renewed further away from the osteoclast source, in \oca s that would otherwise undergo apoptosis. It can thus be expected that a nuclei renewal mechanism may influence, on average, the typical trajectory of osteoclasts within the \bmu\ cutting cone.

\begin{figure}
    \hspace{-3mm}
    \begin{tabular}{ll}
    {\hspace{3mm}\small (a) \m2} & {\hspace{-2mm}\small (b) \m3}
    \\
    \includegraphics[height=0.52\columnwidth]{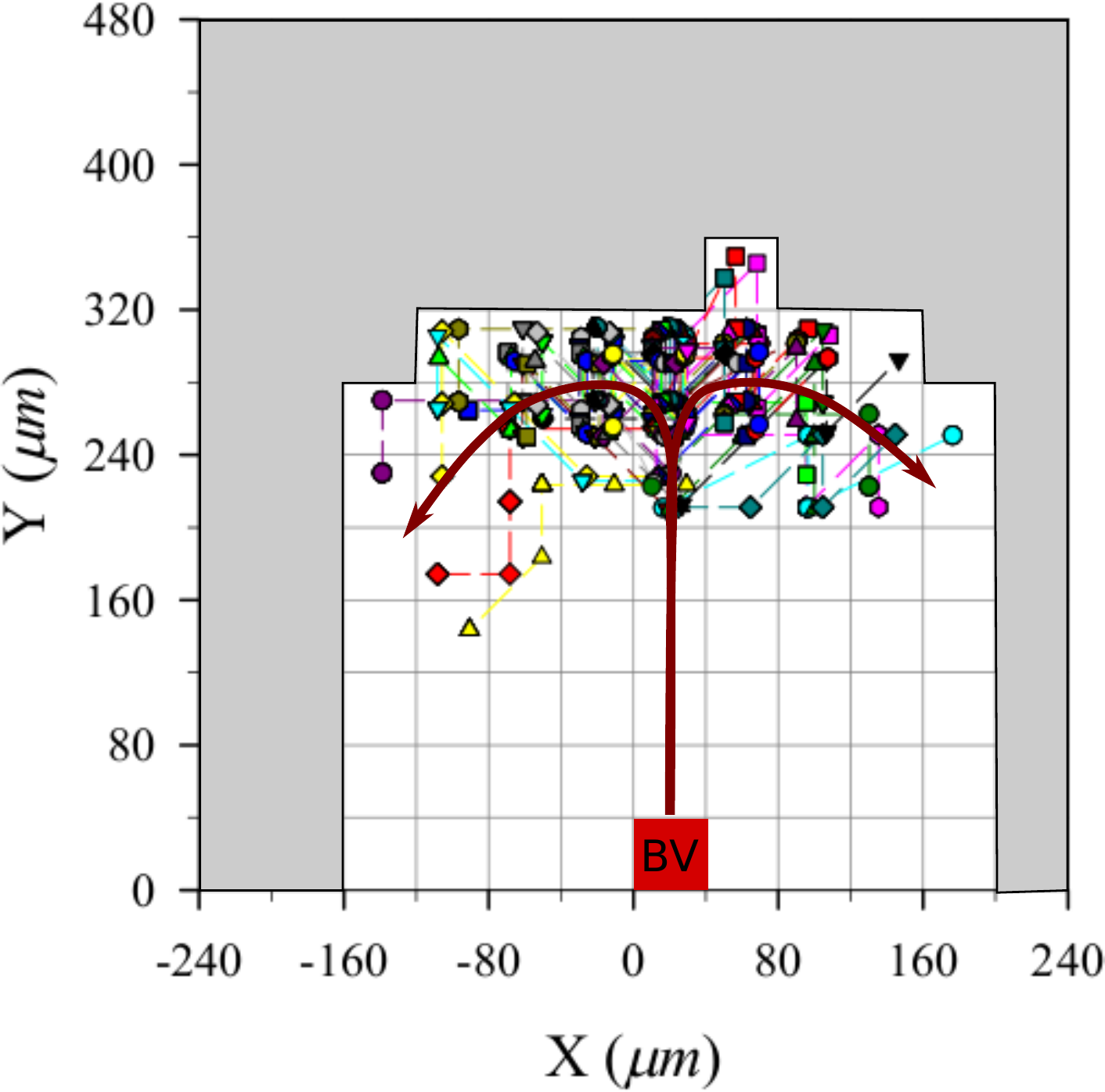} & \hspace{-4mm}\includegraphics[height=0.52\columnwidth]{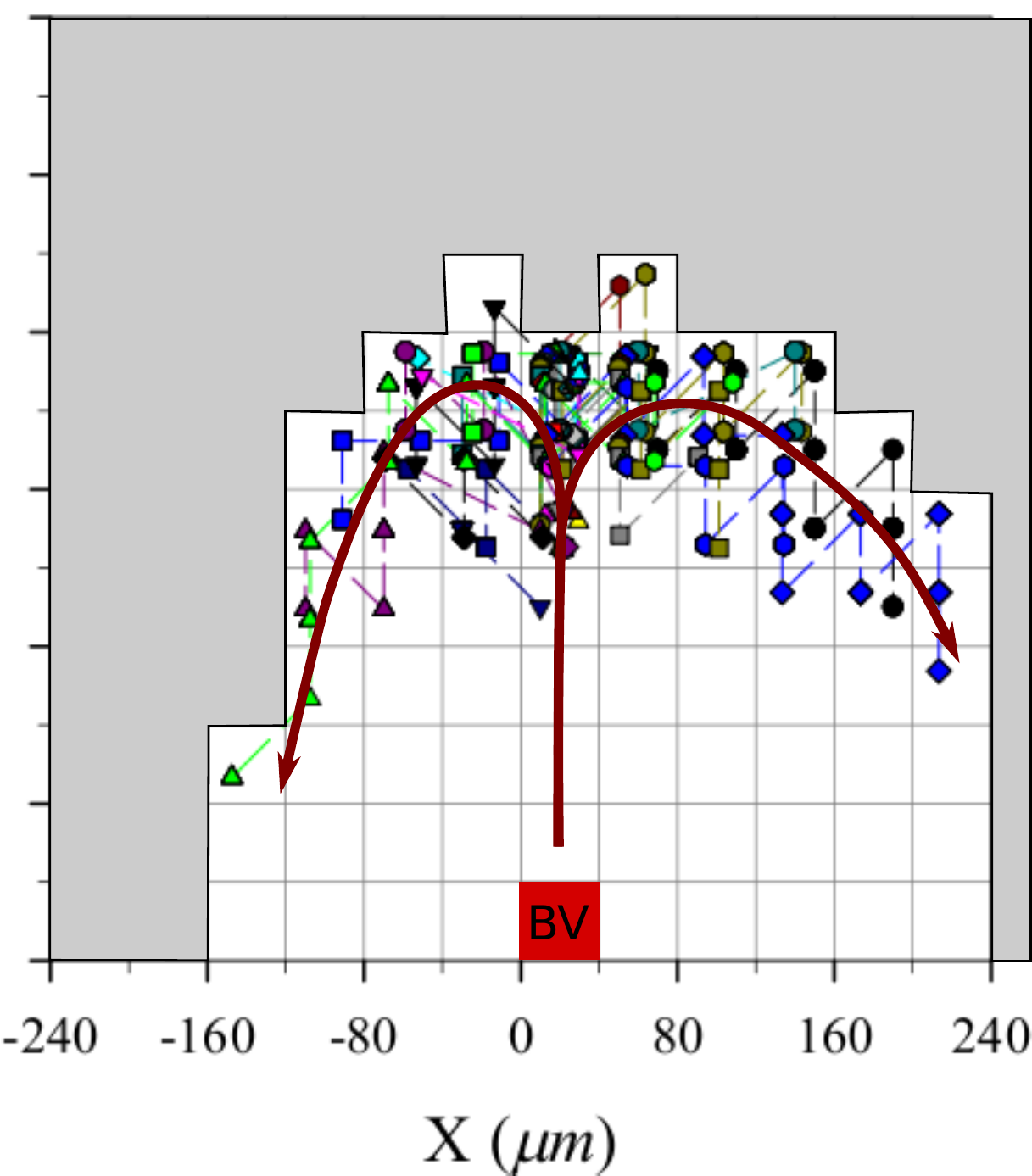}
    \end{tabular}
    \caption{Typical relative trajectories of individual osteoclasts in a reference frame co-moving with the tip of the blood vessel. (a)~\m2; and (b)~\m3. The site occupied by the tip of the blood vessel is shown as well as red arrow lines representing the `treadmill' movement pattern of osteoclasts.}
    \label{Fig:Trajectory}
\end{figure}

The question of the specific trajectories taken by osteoclasts within a \bmu\ cutting cone has been raised by \cite{Burger2003}. \cite{Burger2003} hypothesised that osteoclasts may follow a `treadmill' movement pattern (i.e. following a trajectory from the tip of the blood vessel, where they are formed, to the front of the \bmu, where they mature, and down the sides of the cavity). In Figure~\ref{Fig:Trajectory}, we show the relative trajectories in the \bmu\ taken by all osteoclasts created during $30~\days$ in \m2 and \m3.\footnote{We have checked that these trajectories are typical and representative for both \m2 and \m3 by running the simulations several times with different `random seeds'.} These trajectories are shown from the point of view of an observer moving with the tip of the blood vessel, and so represent the trajectories of the osteoclasts within the cutting cone. In Figure~\ref{Fig:Trajectory}, each osteoclast is assigned a different symbol and a slight offset within a lattice site to be able to distinguish overlapping paths. In both models, osteoclasts are generated in the middle and progressively move towards the sides of the cavity. Most osteoclasts are found to stay within a layer approximately two lattice sites thick against the cavity wall near the front of the \bmu. However, it is seen that the nuclei renewal mechanism introduced in \m3 modifies the movement pattern of the osteoclasts compared to \m2. The increase in lifespan due to nuclei renewal allows osteoclasts to progressively come further towards the back of the \bmu\ in \m3, suggestive of the `treadmill' movement pattern referred to by \cite{Burger2003}.

\section{Conclusions\label{conclusions}}
Our novel computational approach has allowed the investigation of resorption within a single \bmu. In particular, the study of the movement pattern of osteoclasts in the \bmu\ and the effect of the osteoclasts' life history on their collective resorption behaviour is a challenge that computational modelling is able to shed some light on. Our model shows the importance of osteoclast--bone adhesion and of the growth of a blood vessel. Our model suggest that the `clean-up' of the resorption site by extracellular components following resorption by an osteoclast is unlikely to be rate limiting. The model is able to generate cavity shapes that closely resemble those observed experimentally, and produces osteoclast trajectories within the \bmu\ that are consistent with a previous hypothesis by \cite{Burger2003}. Clearly, our simulations already capture essential features of osteoclast resorption behaviour.

Nevertheless, future models can be improved. For example, representing the three dimensional geometry of the \bmu\  cavity would be expected to lead to more realistic simulations of osteoclast movement within the \bmu\  cavity. In our current model, the growth of the blood vessel influences strongly resorption cavity shape, but it is likely that resorption cavity shape in turn influences the growth of the blood vessel, in a feedback interaction. And finally, the \bmu\  progresses through a stress field in bone matrix, encountering osteocytes and various signalling molecules. Including these features in future models is expected to help explain \bmu\  initiation, the diversity of resorption cavity shapes, changes in direction, branching and \bmu\  termination.

\subsubsection*{Acknowledgements}
The authors would like to thank Colin R.~Dunstan for helpful discussions and comments in the preparation of the manuscript, the late Gregory R.~Mundy for his insightful discussions, as well as one of the anonymous reviewers for his/her remarks on osteoclast biology. Support by the Australian Research Council in the framework of the project \textit{Bone regulation - cell interactions to disease} (project number DP0879466, PP), by the National Cancer Institute (grant number U54CA113007, PTC), and by the National Science Foundation (grant number EPS-0919436, PTC) is gratefully acknowledged.

\begin{appendices}
\section{Model description \label{appx:model}}
The technical details of the model description outlined in Section~\ref{model-description} are given below.

\paragraph{Lattice composition.}
The region around the cutting cone of a cortical \bmu\ is known to be mainly composed of mineralised bone matrix, active osteoclasts, bone precursor cells, a blood vessel, and loose connective tissue stroma filling the space between these components~\citep{parfitt1998,martin-burr-sharkey}. Accordingly, each lattice site of the model carries one of the following: (i) mineralised bone matrix, (ii) an osteoclast, (iii) components of a blood vessel, or (iv) loose connective tissue stroma (see Figure~\ref{Fig:BMU}). Bone precursor cells are not considered explicitly in the present model, but their effect for osteoclastogenesis is implicitly accounted for by the generation of new osteoclasts near the tip of a blood vessel.

\paragraph{Blood vessel.}
The blood vessel (\bv) in a cortical \bmu\ provides the local \bmu\ microenvironment with both nutrients and precursor cells (in particular precursor osteoclasts) that are necessary to sustain the remodelling process \citep{parfitt1998}. The blood vessel is assumed in the model to occupy a width of one lattice site and to grow towards the front of the \bmu\ at a maximum rate $v_\bv$ (in $\um/\da$) provided it has enough cavity space to do so. A minimal vertical distance of 280~\um\ (corresponding to 7 lattice sites) between the tip of the blood vessel and the bone surface \citep{parfitt1998} is always enforced by slowing this rate of growth if necessary.

\begin{figure}
    \hspace{-3.5mm}%
    \begin{tabular}{ll}
    (a) \m1 & (b) \m2 and~III
    \\\includegraphics[width=0.49\columnwidth]{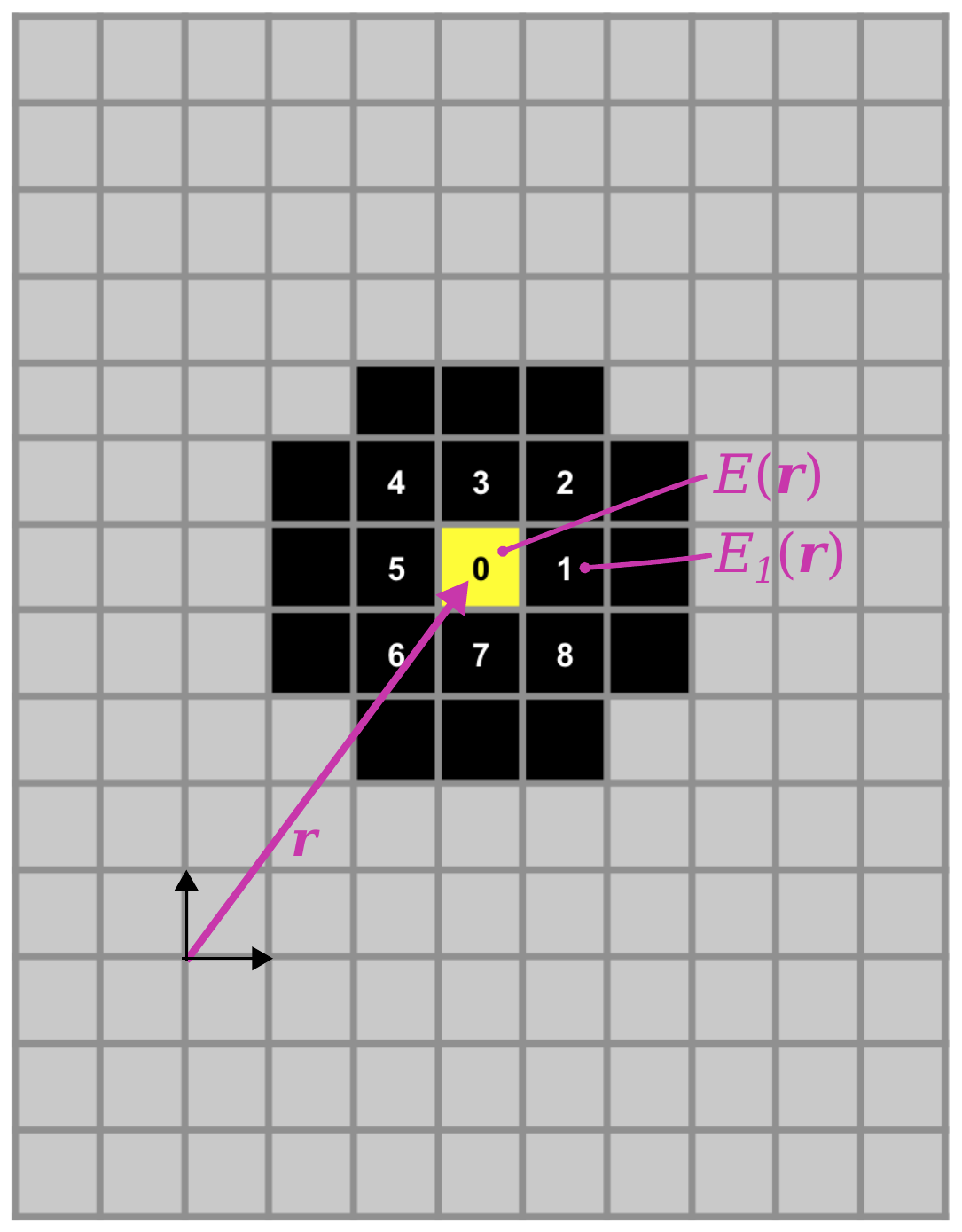}
    &\includegraphics[width=0.49\columnwidth]{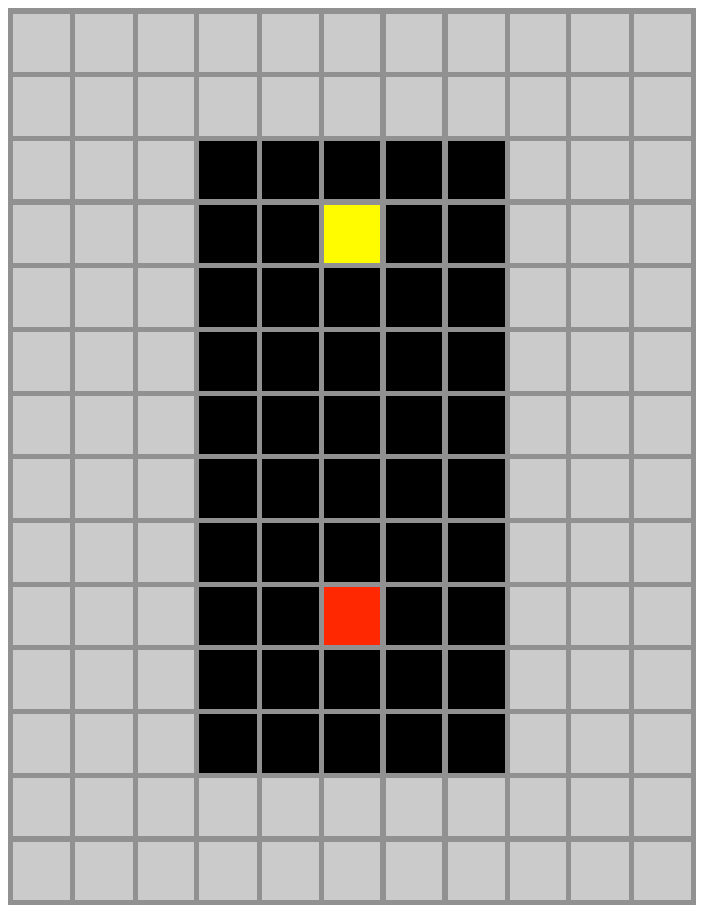}
    \end{tabular}
    \caption{(a) Initial configuration for \m1; (b) initial configuration for \m2 and \m3. In (a), the sites numbered 0--8 denote the nine neighbouring lattice sites of the so-called `Moore neighbourhood' of the \ocm\ (in yellow) located at position $\b r$ on the lattice. These nine lattice sites correspond to the possible locations where the \ocm\ can migrate to during a time increment, including no movement (resting phase). The total energies $E(\b r)$ at $\b r$, and $E_1(\b r)$ at the Moore neighbour `1', are also mentioned.}
    \label{Fig:Neighbor}
\end{figure}

\paragraph{Maturation and activation of osteoclasts.}
The maturation of osteoclasts in a \bmu\ is the result of a cascade of events that take place around the tip of the blood vessel. This cascade of events is known to involve several cell types and molecules (such as pre-osteoblasts, \mcsf\ and \rankl\ etc., see \citep{roodman,martin}). While these cell types and molecules are not explicitly considered in the present model, we assume that their net effect is to generate new mature osteoclasts at a rate $\eta_\oc$ (in units of $\da^{-1}$) at a distance of about $240~\um$ or 6 lattice sites ahead of the tip of the blood vessel.\footnote{The exact position at which newly-generated \ocm s are placed in the model depends on whether a lattice site is free of other osteoclasts. Specifically, this position is chosen to be in a range of $160$--$240~\um$ ahead of the tip of the blood vessel \citep{parfitt1998}, corresponding to 4--6 lattice sites. New \ocm s are thus placed at a distance of $40$--$120~\um$  ($1$--$3$ lattice sites) from the front of the cavity surface. In all the simulations performed, this range always allowed the positioning of new \ocm s on osteoclast-free stromal sites, ensuring a constant generation rate $\eta_\oc$ at all times.} These \emph{mature osteoclasts} (which we  denote by \ocm s) are assumed to be fully differentiated:~they represent multinucleated cells capable of migrating through the connective tissue stroma and of attaching to the bone surface for its resorption. Mature osteoclasts that are attached to the bone surface and are actively resorbing the bone matrix are referred to as \emph{active osteoclasts} (which we denote by \oca s). The attachment of an osteoclast to the bone surface and its activation is in reality a complex process that requires several signalling pathways to be activated, in particular the \rank-\rankl\ pathway \citep{roodman,martin}. In our model, it is considered that a mature osteoclast becomes `active' as soon as it is adjacent to or diagonal with a bone lattice site (i.e., as soon as it is in the so-called `Moore neighbourhood' of a bone lattice site, see Figure~\ref{Fig:Neighbor}(a)). An active osteoclast \oca\ is assumed to stay active until all the bone sites in its Moore neighbourhood are resorbed, after which the \oca\ becomes a migrating \ocm\ again \citep{fukushima-bekker-gay}.

\paragraph{Migration behaviour of osteoclasts---evolution rules.}
Several mathematical approaches exist in the literature to represent the migration of cells in their microenvironment \citep{noble,schweitzer,peruani-morelli,othmer-etal}. Here, we represent the migration of \ocm s through connective tissue stroma as a biased random walk \citep{vankampen,schweitzer,othmer-stevens}. The migration behaviour of an \ocm\ is specified in the model by so-called `evolution rules'. The evolution rules determine towards which neighbouring site the \ocm\ will migrate during the next time increment. This site is chosen randomly in the Moore neighbourhood of the \ocm\ with unequal probabilities that depend on the lattice composition in the \ocm's surroundings. The immediate surroundings of an osteoclast predominantly determine the cell's migration through the extracellular connective tissue stroma, or the cell's adhesion to other cells or the bone surface, via a number of local biochemical signals mediated by enzymes, cytokines, cell adhesion molecules etc. However, to an individual cell, these biochemical signals have a fluctuating character. Fluctuations are due to small-scale inhomogeneities in the connective tissue, or to fluctuations in signalling pathways or in biochemical reactions occurring at multiple levels; from the molecular level (e.g.\ thermal fluctuations) to the cellular level (e.g.\ membrane fluctuations) to the organism level (e.g.\ food intake and other exogeneous fluctuations) \citep{schienbein-etal,mombach-glazier}. The probabilistic character of the migration behaviour of the \ocm s precisely accounts for such fluctuations in local biochemical signals. The total interaction energy $E(\b r)$ between an \ocm\ located at position $\b r$ on the lattice and entities located in the Moore neighbourhood of the \ocm\ determines the strength of the biases introduced in the random walk.

In \citep{beysens-etal,drasdo-hoehme,block-schoell-draso}, the strength of the above-mentioned background of metabolic fluctuations is quantified by the so-called `metabolic energy' $F_T$. Depending on the strength of the metabolic energy $F_T$ relative to the total interaction energy $E(\b r)$, the migration behaviour of the \ocm\ is given more or less randomness, much in the same way that thermal energy gives more or less random fluctuations in `Brownian motion' of atoms or molecules at the molecular level. The probability for an \ocm\ at $\b r$ to migrate to its Moore neighbour $i$ (Figure~\ref{Fig:Neighbor}(a)) is assumed to be given by
\begin{align}\label{jump-probabilities}
    P_{i}(\b r) = \frac{\exp(-E_{i}(\b r)/F_T)}{\sum\limits_{j=0}^{8} \exp(-E_{j}(\b r)/F_T)},
\end{align}
where $E_i(\b r)$ denotes the total energy at the Moore neighbour $i$. The migration scheme~\eqref{jump-probabilities} is such that the cell is more likely to migrate towards neighbouring lattice sites $i$ at which the \ocm's total energy $E_i(\b r)$ is minimum. The denominator ensures that $\sum_{i=0}^8 P_i(\b r) = 1$. If the metabolic energy is high ($F_T \gg |E_i(\b r)|$), biases in the probabilities $P_i(\b r)$ are reduced, leading to more erratic migration. If the metabolic energy is low ($F_T \ll |E_i(\b r)|$, biases in the probabilities $P_i(\b r)$ are accentuated, leading to more persistent migration towards the local minimum of energy.

Schemes similar to Eq.~\eqref{jump-probabilities} have been previously used for cellular automata in the literature (see, \textit{e.g.}, \citep{block-schoell-draso,ghaemi-shahrokhi}). The scheme~\eqref{jump-probabilities} essentially differs from the traditional Glauber dynamical scheme and the Metropolis dynamical scheme \citep{Glauber1963} in that the system does not evolve towards equilibrium of the metabolic energy $F_T$ (corresponding to thermal equilibrium if $F_T=k_\text{B} T$).\footnote{Indeed, the scheme~\eqref{jump-probabilities} does not satisfy detailed balance of the corresponding Master equation with Boltzmann-like stationary distribution \citep{vankampen}.} This is not contradictory, as a cortical \bmu\ is a nonequilibrium entity, requiring a continual source of new cells and metabolic energy to maintain itself.

\paragraph{Interaction energies.} The total energy $E(\b r)$ of an \ocm\ at $\b r$ on the lattice is the sum of the interaction energy between the \ocm\ and other osteoclasts, the interaction energy between the \ocm\ and bone, and the interaction energy between the \ocm\ and blood vessel components. We denote these interaction energies by $E_j^{\oc-\oc}$, $E_j^{\oc-\bone}$ and $E_j^{\oc-\bv}$, respectively. The subscript $j$ indicates the relative position of the entity that the \ocm\ interacts with. The range of these interaction energies is assumed to be restricted to the Moore neighbourhood of the \ocm, so $j$ represents a Moore neighbour index and the total energy is the sum:
\begin{align}\label{potential-energy}
       E(\b r) = \sum_{j=0}^{8} &\Big\{N_j^\oc E^{\oc-\oc}_j + N_j^\bone E^{\oc-\bone}_j \notag
       \\&+ N_j^\bv E^{\oc-\bv}_j\Big\},
\end{align}
where $N^\alpha_j = N^\alpha_J(\b r)$ is equal to $1$ if the Moore neighbour $j$ is occupied by the entity $\alpha$, and equal to $0$ otherwise ($\alpha = \oc,\bone,\bv$). We define $E_j^{\oc-\oc}$, $E_j^{\oc-\bone}$ and $E_j^{\oc-\bv}$ by
\begin{align}
    E^{\oc-\oc}_j &= \begin{cases}
        E_{\oc-\oc}^\text{fusion}, &\quad j=0,
    \\ E_{\oc-\oc} < 0, &\quad\forall j=1,...,8,
    \\ 0 &\quad j \not\in \text{Moore neighbour},
    \end{cases}\label{energies-vs-distance-oc}
    \\E^{\oc-\bone}_j &= \begin{cases}
        +\infty, &\quad j=0,
    \\ E_{\oc-\bone} < 0, &\quad\forall j=1,...,8,
    \\ 0 &\quad j \not\in \text{Moore neighbour},
    \end{cases}\label{energies-vs-distance-bone}
    \\E^{\oc-\bv}_j &= \begin{cases}
        +\infty, &\quad j=0,
    \\ 0, &\quad\forall j\neq 0,
    \end{cases}\label{energies-vs-distance-bv}
\end{align}
where $j=0,...,8$ indexes the Moore neighbours of the \ocm. In Eq.~\ref{energies-vs-distance-oc}--\ref{energies-vs-distance-bv}, the energy value at $j=0$ represents an `exclusion energy'. If infinite, this exclusion energy corresponds to a so-called `hard-core repulsion'. If finite, the \ocm\ may coexist with another entity on the same lattice site. Energy values at $j=1, ..., 8$ determine short-range interaction between the \ocm\ and an entity on a neighbouring site. These energy values determine the adhesion properties of the \ocm\ with its neighbouring entities. Consequently:
\begin{itemize}
\item Eq.~\eqref{energies-vs-distance-oc} specifies that if $E^\text{fusion}_{\oc-\oc}$ is finite ($<+\infty$), an \ocm\ may migrate onto a site previously occupied by another osteoclast, in which case fusion with this existing osteoclast is assumed to take place (see also `nuclei renewal' below). Because $E_{\oc-\oc}$ is negative, an osteoclast--osteoclast adhesion is assumed;

\item Eq.~\eqref{energies-vs-distance-bone} specifies that an \ocm\ cannot occupy the same lattice site as bone. Because $E_{\oc-\bone}$ is negative, an osteoclast--bone adhesion is assumed;\footnote{The assumption that the interaction with an orthogonal site ($j=2,4,6,8$) has the same strength as that with a diagonal site ($j=1,3,5,7$) in Eqs.~\eqref{energies-vs-distance-oc} and~\eqref{energies-vs-distance-bone} implicitly introduces a lattice anisotropy. Such lattice anisotropies are unavoidable in any regular lattice \citep{markus-hess,drasdo-2005,nishiyama-tokihiro}. Still, the Moore neighbourhood is expected to limit lattice artefacts compared to the von Neumann neighbourhood \citep{Poplawski2007} and it allows in our model to maximise the interaction of an osteoclast with bone sites.\label{fn:lattice-artefacts}}

\item Eq.~\eqref{energies-vs-distance-bv} specifies that an \ocm\ cannot occupy the same lattice site as blood vessel components, and that an \ocm\ shows no particular preference to adhere to blood vessel components.
\end{itemize}

The concept of `interaction energies' is a high-level simplification for complex molecular processes, but it allows to integrate the underlying biochemical signals and resultant cell properties into a single concept. In \citep{beysens-etal,drasdo-hoehme,block-schoell-draso}, it is estimated that such biochemically-induced interaction energies are typically of the order $1\,F_T$--$10\,F_T$. In this paper, all energies are measured in units of the metabolic energy $F_T$ and their physiological range is assumed to be within $0~F_T$--$10~F_T$.\footnote{\label{footnote:energy-range}This narrow range of values restricts the allowable parameter space substantially.}

\paragraph{Migration behaviour examples.}
We provide two examples of the migration behaviour of an \ocm\ in response to its surrounding. (i)~When an \ocm's microenvironment is composed of connective tissue stroma only, all neighbouring lattice sites have the same total interaction energy and the probability to migrate to any of those sites is equal: the \ocm's\ migration resembles a so-called `random walk' \citep{vankampen,schweitzer,othmer-stevens} and results in an isotropic diffusive motion for the \ocm, with an effective diffusion coefficient $D_\oc$ (see below). (ii)~For the situation depicted in Figure~\ref{Fig:Neighbor}(a), the \ocm\ located on the lattice site `0' is more likely to migrate to even-numbered sites than to odd-numbered sites. Indeed, on even-numbered sites, the presence of a neighbouring bone site lowers the energy due to a negative osteoclast--bone adhesion energy $E_{\oc-\bone}$.

\paragraph{Osteoclast lifespan: nuclei renewal, cell apoptosis.}
In our model, \ocm s that are newly generated near the tip of the blood vessel are initially assigned a fixed lifespan $\tau_\oc$. To account for a lifespan-increasing nuclei renewal process in our simulations, we consider that migrating \ocm s can fuse with existing \oca s or \ocm s with different probabilities $E_{\ocm-\oca}^\text{fusion}$ and $E_{\ocm-\ocm}^\text{fusion}$, respectively (see Eq.~\eqref{energies-vs-distance-oc}). The lifespan of the osteoclast resulting from this fusion is increased by the remaining lifetime of the fusing \ocm, and the latter \ocm\ is removed from the system. This fusion process is assumed to refresh at once the nuclei of the resulting osteoclast. The resulting osteoclast is not assumed to grow any bigger as a result of nuclei fusion. It is implicitly assumed that older nuclei of the cell are degraded and removed when new nuclei are added. This fusion algorithm between two multinucleated cells can be interpreted to represent multiple fusions by mononuclear cells jointly with individual nuclei degradation.

Depending on whether the fusion energy $E_{\oc-\oc}^\text{fusion}$ is greater or lesser than the adhesion energy $E_{\oc-\oc}$, fusion or adhesion is preferentially selected by the stochastic scheme~\eqref{jump-probabilities}--\eqref{energies-vs-distance-oc}. According to the observations by \cite{fukushima-bekker-gay}, we assume that an \ocm\ and an \oca\ should preferentially fuse, while two \ocm s should preferentially not fuse. We thus consider that
\begin{align}
    &E_{\ocm-\oca}^\text{fusion} \ll E_{\oc-\oc} \ll E_{\ocm-\ocm}^\text{fusion},
\end{align}
and take $E_{\ocm-\ocm}^\text{fusion}=+\infty$ (no fusion between two \ocm s). Note that when an \ocm\ has several nearby osteoclasts that it could fuse with, the algorithm selects preferentially the ones that are active and next to the maximum number of bone sites. In case of osteoclasts with identical properties, the probability to fuse with any one of them is equal.

When the age of an osteoclast reaches its alloted lifespan (whether that lifespan has been increased by nuclei renewal or not), the cell is assumed to undergo apoptosis irrespective from its activity state (migrating or resorbing) and it is removed from the system. We assume that the renewal of an osteoclast's nuclei does not influence that osteoclast's activity state \citep{miller,fukushima-bekker-gay}.

\paragraph{Bone matrix dissolution.}
An \oca\ resorbs a single neighbouring bone site at a time, in such a way that the density of this bone site decreases exponentially with time:
\begin{align}
	\frac{\text{d}m(t)}{\text{d}t}=-\gamma m(t),\quad \text{or}\quad m(t+\Delta t)=m(t)\text{e}^{-\gamma \Delta t},
\label{bone-density-decay}
\end{align}
where $m(t)\in [0,1]$ denotes the relative bone density (normalised by fully-mineralised bone density) and $\gamma=10/\da$ is the dissolution rate. The kinetic law~\eqref{bone-density-decay} for dissolution, which prescribes a rate of bone density loss proportional to bone density, is motivated by chemical bone dissolution kinetics, where the efficiency of the dissolution changes with the mineral component composition and density \citep{boyde-jones,grynpas-cheng,huang-etal2006}. Bone matrix with density lower than a critical bone density value $m^\ast=0.1$ is assumed to become a cavity site filled with connective tissue stroma.

To account for a possible rate-limiting extracellular collagen digestion, we introduce in our model a hypothetical inhibition period $\tauclean$ during which a newly exposed bone surface (due to resorption) is inaccessible to \ocm s. This is achieved within the scheme~\eqref{jump-probabilities} by tagging newly-resorbed sites momentarily (before they become stromal sites) as ``quiescent'' bone sites: an \ocm\ in the proximity of such a site is prevented to migrate through it and is prevented to become activated into an \oca, unless other bone sites are in the \ocm's neighbourhood.

\paragraph{Initial conditions and simulation parameters.}
The initial lattice configurations used in \m1, \m2 and \m3 are shown in Figure~\ref{Fig:Neighbor}. The evolution of the system is simulated for 30 days with time increments of $\Delta t=0.1~\da$. At each time increment, all osteoclasts are selected in random order and an update of their state and position is performed according to the flow chart in Figure~\ref{Fig:FlowChart} (i.e., asynchronous update). The dissolution of bone sites by the \oca s is implemented in this selected order too. The possible growth of the blood vessel as well as the generation of new \ocm s is performed last.

Finally, it is noted that the time increment $\Delta t$ cannot be chosen arbitrarily. Indeed, $\Delta t$ and the lattice step size $\sigma$ determine the diffusion coefficient $D_\oc$ for the pure random walk (in stromal tissue only) of the \ocm s. With the Moore neighbourhood, one has $D_\oc = \sigma^2/(3\Delta t) \approx 6.2\cdot 10^{-10}~\cm^2/\second$.\footnote{For random walks in two dimensions, the mean square displacement is $\langle R^2(t)\rangle = 4 D_\oc t$, where $D_\oc$ is the diffusion coefficient. This mean square displacement can also be formulated as the product of the mean number of time steps $\langle k\rangle$ required to observe a jump with nonzero length and the mean square displacement $\langle \ell^2 \rangle$ during such a jump. Considering that the 2D Moore neighbourhood includes the possibility of resting, jumps of nonzero length only occur $8/9^\text{th}$ of the time, so $\langle k\rangle=(8/9)t/\Delta t$. The average length of a jump with nonzero length is $\langle \ell^{2}\rangle = \tfrac{1}{8}[4\cdot \sigma^2 + 4\cdot(\sigma \sqrt{2})^2]=(3/2)\sigma^{2}$. Thus, $D_{OC} = \langle R^{2}(t)\rangle/(4t) = {(1/4)}\cdot {(3/2)\sigma^{2}} \cdot {(8/9)/\Delta t} = \sigma^2/(3 \Delta t)$.}

In Table~\ref{Table:Parameter} we list all the parameters of the model along with the range of values investigated in Section~\ref{results}, and a so-called `default' value, which is assigned unless that parameter is explicitly varied. These ranges of values implicitly exclude regions of the parameter space that were either not physiological, or that were leading to physiologically unrealistic results (e.g. an abnormally-large resorption cavity or a too slow \bmu\ progression rate). Default values are shown in bold, and do lead to realistic resorption cavity geometries (see Section~\ref{discussion}).

\begin{table*}[bh!]
\caption{List of model parameters}
\label{Table:Parameter}
\centering\begin{tabular}{l l l}
\hline
Parameters & Description & Values (\bf default) \\
\hline
$\sigma$		& lattice step size					& $40\, \um$ \\
$D_\oc$			& diffusion coefficient of an osteoclast (pure random walk)	& $\num{0.62e-10}~\cm^2/\second$\\ 
$E_{\oc-\oc}$	& osteoclast--osteoclast adhesion energy			& $\lbrack -4, 0\rbrack$ ($\b{-1}$) $F_T$\\
$E_{\oc-\bone}$	& osteoclast--bone adhesion energy					& $\lbrack -4, 0\rbrack$ ($\b{-4}$) $F_T$\\
$\rule[-1.2ex]{0pt}{3.6ex}E_{\ocm-\oca}^\text{fusion}$& fusion energy between a mature osteoclast and an active osteoclast & $\lbrack -8, 0\rbrack$ ($\b{-4}$) $F_T$\\
$\rule[-1.2ex]{0pt}{3.6ex}E_{\ocm-\ocm}^\text{fusion}$& fusion energy between mature osteoclasts & $+\infty$ $F_T$\\
$m_0$		& initial relative bone density at all bone sites			& $1$ \\
$m^{*}$			& critical relative bone density under which a bone site is removed							& $0.1$ \\
$\gamma$		& dissolution rate of relative bone density by active osteoclasts						& $10/\da$ \\
$\tau_\oc$		& initial lifespan of osteoclasts						& $\b{2}, 4, 6, 8~\days$ \\
$\eta_\oc$		& generation rate of osteoclasts by the blood vessel 				& $1.25, \b{1.66}, 2.5, 5.0\, \da^{-1}$ \\
$v_\bv$			& maximum rate of growth of the blood vessel						& $4, 8, 20, \b{40}\, \um / \da$ \\
$\tauclean$		& period of inhibition of further resorption at a newly-resorbed bone site& $\b{0}, 1, 2, 4, 5, 8, 10~\days$ \\
\hline
\end{tabular}
\end{table*}
\end{appendices}

\clearpage 
\bibliographystyle{apalike} 
\bibliography{CA}
\end{document}